\documentclass[amsmath,amssymb,prl,twocolumn,superscriptaddress]{revtex4-2}
\usepackage{physics}
\usepackage{graphicx}
\usepackage{bm}
\usepackage{color}

\newcommand{\MM}[1]{{\color{magenta}#1}}

\begin{document}
\title{Photo-induced sliding transition into a hidden phase in van der Waals materials}
\author{Jiajun Li}
\affiliation{Laboratory for Theoretical and Computational Physics, Paul Scherrer Institute, 5232 PSI Villigen, Switzerland}
\affiliation{Department of Physics, University of Fribourg, 1700 Fribourg, Switzerland}
\author{Philipp Werner}
\affiliation{Department of Physics, University of Fribourg, 1700 Fribourg, Switzerland}
\author{Michael A. Sentef}
\affiliation{Max Planck Institute for the Structure and Dynamics of Matter, Center for Free-Electron Laser Science (CFEL), Luruper Chaussee 149, 22761 Hamburg, Germany}
\author{Markus M\"{u}ller}
\affiliation{Laboratory for Theoretical and Computational Physics, Paul Scherrer Institute, 5232 PSI Villigen, Switzerland}
\date{\today}
\begin{abstract}
We propose a generic scenario for metastability and excitation-induced switching in layered materials. Focusing on a minimal bilayer stack, where each layer consists of a honeycomb lattice with A and B sublattices, we map out the energy landscape with respect to the relative sliding of the layers. The sliding affects the interlayer hopping, which induces a splitting between bonding and anti-bonding bands. When this splitting is large, the AA and AB stacking configurations correspond to the global and secondary minima, respectively, and these configurations are separated by a barrier against layer-sliding. While chemical doping only flattens this barrier, strong \emph{photodoping} from bonding to antibonding bands can transiently destabilize the global minimum and induce a sliding motion toward the AB stacked configuration, thereby switching from an equilibrium insulator to a nearly gapless metastable phase. This hopping-driven effect is enhanced by local repulsive interactions, which increase the gap and facilitate the inter-layer sliding. 
\end{abstract}
\maketitle

The engineering of on-demand properties of quantum materials via ultrafast laser excitation is a rapidly growing field \cite{basov2017,delatorre2021}. One promising direction is the exploration of nonthermal pathways to control the crystal lattice and thereby the electronic properties of a given material. Layered van der Waals materials are an ideal platform for such explorations. The weak bonding between internally rigid layers opens opportunities to design and manipulate these systems \cite{geim2013,novoselov2016}. Since the first fabrication of graphene, the art of stacking atomically thin layers into precisely controlled heterostructures has developed into a routine technology for exploring new 
material properties that are very sensitive to the nature of the stacking. A prominent example are Moir\'{e} superlattices, that can be used as tunable quantum simulators \cite{cao2018, cao2018correl, kennes2021}. 

Due to the weak interlayer correlations, the individual rigid  layers can slide relative to each other when the energy landscape is modulated, as has recently been observed in WTe$_2$ \cite{fei2018} and hBN multilayers \cite{yasuda2021,vizner_stern2021}. Certain stacking configurations result in an electric dipole moment whose coupling to a strong electric field allows for controllable sliding into an otherwise metastable stacking configuration \cite{li2017,wu2021,tang2022}. 

The effect of the stacking arrangement on the electronic structure of the low-temperature phase of 1$T$-TaS$_2$ has also been the subject of numerous recent investigations \cite{nicholson2022,wu2022, lee2019, petocchi2022}. In this correlated material, the stacking pattern of the surface layers can be altered  
on ultrafast timescales using either laser \cite{stojchevska2014,stahl2020,maklar2022} or strong electric-field pulses \cite{hollander2015,ma2016,cho2016,vaskivskyi2016,ravnik2021}. These nonequilibrium manipulations result in a metallic hidden phase which seems inaccessible under equilibrium conditions \cite{stojchevska2014}. 
While the strong pulses excite phonons, they also modulate the electronic kinetic energy through photodoping \cite{ligges2018,maklar2022}, and thus the precise nonthermal pathway toward the hidden stacking pattern has remained unclear.

Inspired by such experiments, we propose here a concrete and  generic scenario in which a substantial photoinduced transfer of electrons to unoccupied antibonding bands induces interlayer sliding toward an initially metastable state in a bilayer material, realizing a 
solid-state analogon of photo-induced conformation changes in molecules. Our aim is two-fold: We introduce a minimal tight-binding model whose electronic kinetic energy exhibits an asymmetric double-well structure (i.e., metastability) in the continuous space of stacking configurations. We further show that photodoping can destabilize the equilibrium stacking and induce spontaneous sliding into the secondary energy minimum,  amounting to light-induced stacking shifts. 

\begin{figure*}
\includegraphics[scale=0.7]{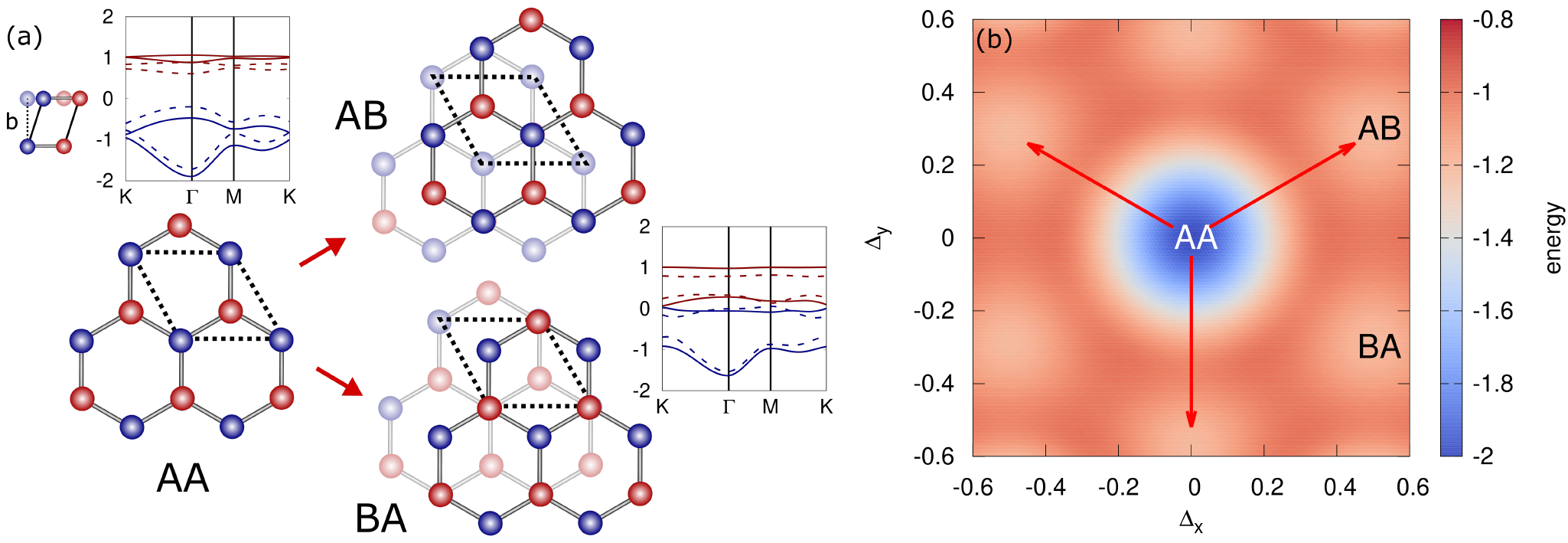}
\caption{Kinetic energy $E(\bm\Delta)$ for different stackings of a bilayer of honeycomb lattices. (a) Two likely sliding directions. A unit cell (of side $a=1$) is delineated by dotted lines. For identical atoms A and B, AB and BA stackings are equivalent. The four bands for AA (left, $\bm \Delta = (0,0)$) and AB stacking (right, $\bm\Delta=(1/2,\sqrt{3}/6)$) 
are shown as solid lines. The dashed lines correspond to intermediate shifts $\bm\Delta=(1/5,0)$ (and $(7/10,\sqrt{3}/6)$), that increase the shortest interlayer atomic distances, as illustrated in the
sideview  (top left).
(b) 
$E(\bm\Delta)$
as a function of the sliding vector $\bm \Delta$, for parameters $t_1=10t_0,\xi=0.6b$.
Six sliding directions lead from the AA-stacked global minimum to  AB/BA-stacked local minima.}
\label{stacking}
\end{figure*}

\emph{Model--} 
We consider materials consisting of rigid layers that can slide relative to each other. We assume that the associated variation of the hopping energy dominates over that of the van der Waals attraction, which is usually weak. We consider two parallel (untwisted) rigid lattices separated by a fixed distance $b$, and characterize the stacking of the layers by a relative shift vector $\bm \Delta$. We describe the coupled layers by a single-orbital electronic Hamiltonian,
\begin{align}
H[\bm\Delta]=H_0[\bm\Delta] + U\sum_{i\alpha}(n_{i\alpha\uparrow}-1/2) (n_{i\alpha\downarrow}-1/2),
\end{align}
where $H_0=-\sum_{ij \alpha\alpha'\sigma}t_{ij;\alpha\alpha'}(\bm \Delta)c^\dag_{i\alpha\sigma} c_{j\alpha'\sigma}$ is the kinetic energy,
with $c^\dag_{i\alpha\sigma}$ the creation operator of an electron at site $i$ in the layer $\alpha=1,2$, while $\sigma=\uparrow,\downarrow$ labels the spin degree of freedom. $U$ is an onsite Coulomb repulsion, and $n_{i\alpha \sigma} = c^\dag_{i\alpha}c_{i\alpha}$. We assume spin degeneracy and thus omit the spin index for simplicity. The hopping amplitudes $t_{i j;\alpha\alpha'}$ are assumed to decay exponentially over a scale $\xi$, with different prefactors for intra- and interlayer hopping, $t^{\text{intra}}_{ij}=t_0\exp((d_{ij;0}-d_{\text{min};0})/\xi)$ and  $t^{\text{inter}}_{ij}=t_1\exp((d_{ij;1}-d_{\text{min};1})/\xi)$
 (see supplemental material, SM), with intra- (inter-)layer distances $d_{ij;0}$ ($d_{ij;1}$) and minimal distances $d_{\text{min};0}$ ($d_{\text{min};1}$), respectively. $t^{\text{inter}}_{ij}$ is maximal ($=t_1$) when the atoms $i,j$ are stacked on top of each other.

\emph{Kinetic metastability -- }As a minimal model to study the hidden phase of a bilayer system, we consider a honeycomb lattice with two atoms (A and B, blue and red in Fig.~\ref{stacking}) per unit cell and one electron per atom. We assume A and B to be  identical atoms, unlike in interfacial ferroelectrics, although our analysis can be easily generalized. The lattice constant (side of the unit cell in Fig. 1) is $a=1$, and we assume $b=0.5a$\MM{~\footnote{A small $b$ captures realistic layers with a large super-unit cell}}. 
The four bands of the noninteracting model are shown in Fig.~\ref{stacking}(a) for AA  and AB stacking, exhibiting two or only one pair of strong bonding/antibonding bands, respectively. 
In equilibrium, the lower two bands are filled. The energy landscape $E(\bm \Delta)=\langle H_e(\bm\Delta)\rangle$ evaluated for $U=0$ is plotted in Fig.~\ref{stacking}(b). Strong bonding bands make AA stacking the global minimum, while AB/BA stacking corresponds to a local minimum. This is generically the case for stackings with fewer pairs of strongly bonding atoms 
(see SM). Such local minima are often metallic (or tiny-gap  semiconductors as in the present model), whereas the AA minimum is  insulating, as seen from the inset of Fig.~\ref{exy}(a). 
This is ensured by a strong interlayer hopping 
$t_1 \gg t_0$ 
which separates the bonding and antibonding bands by a substantial hybridization gap. When $\bm\Delta$ deviates from ${(0,0)}$, the distance between the most closely stacked atoms increases.
Accordingly, the interlayer hybridization decreases, the gap shrinks and the total energy increases. At the AB minimum, the high-energy bonding and antibonding bands remain widely separated, while the other two bands approach each other and barely avoid touching in two Dirac points,
see the right inset of Fig.~\ref{stacking}(a). 

\begin{figure}
\includegraphics[scale=0.72]{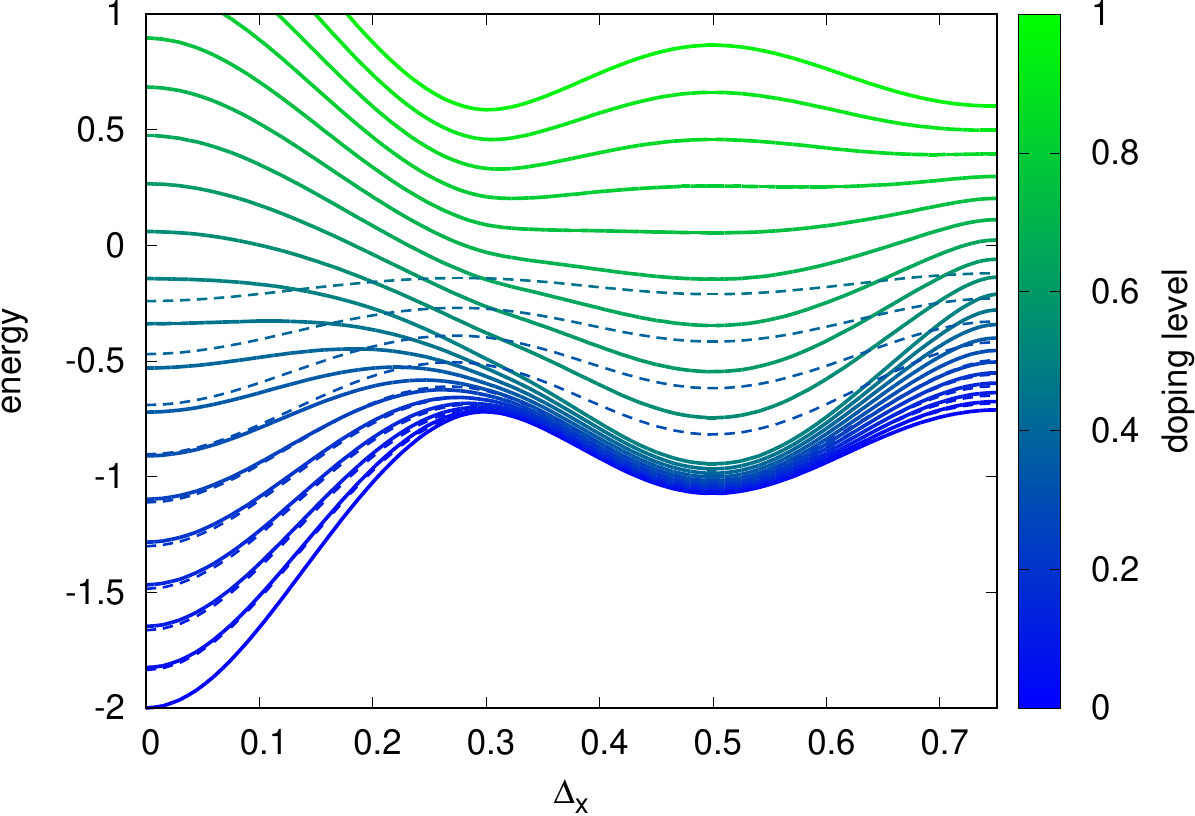}
\caption{Photodoping vs. chemical doping. Increasing doping (from blue to green) of either type  reduces the energetic difference between the two minima, but only photodoping (solid lines) is capable of destabilizing the global minimum at AA stacking. For photodoping, the doping level $f$ is the fraction of electrons moved from the lower two bands to the upper two bands, and increases by steps of 0.05 between consecutive solid curves. For chemical doping,
the hole-doping level $(1-n)/2$ also increases in steps of 0.05 (dashed lines). A large interlayer hopping $t_1=40t_0$ and $\xi=0.4b$ was chosen to  demonstrate the effect clearly.  }
\label{fn}
\end{figure}

\emph{Destabilization by photodoping --} We now show that photodoping can induce a transition from the global to a local minimum. 
We restrict ourselves to sliding along the high-symmetry direction $\Delta= \Delta_x(1,\sqrt{3}/3)$ (red arrow in Fig.~\ref{stacking}(b)) from AA to AB ($\Delta_x=1/2$). 

Careful photo-excitation protocols in few-layer materials may create an inverted population among the bands near the chemical potential \cite{li2018}, and negative temperature states in extreme cases \cite{tsuji2011}. Provided that the driving ensures that all bands are occupied close to their bottom, this destabilizes the global minimum (see SM {for a general argument in the case of full population inversion}) 
since the energy of the now occupied antibonding bands (red in the insets of Fig.~\ref{stacking}) decreases with the band gap as the stacking deviates from AA. The electrons within each band typically thermalize within tens of femtoseconds  to an electronic temperature through electron-electron scattering, and cool down within picoseconds to the temperature of the crystal lattice through electron-phonon scattering.
In contrast, the inter-band relaxation
of a heavily inverted population 
can take much longer if a large energy gap (exceeding the range of phonon energies) separates the population-inverted bands, and if Auger recombinations are slow as well~\footnote{This may arise either due to a gap exceeding the relevant electronic bandwidths, or more generally due to the weakness of  Coulomb-mediated Auger processes that involve relatively large momentum transfers.}, leaving only very slow radiative recombination, or high-order phonon scattering. This suggests to model the photoinduced electronic distribution after intra-band relaxations by two separate chemical potentials, above and below the central gap.

In the following, we assume that 
photodoping has transferred a fraction $f$ of the valence electrons to the lowest conduction band states
without creating significant entropy,  thus assuming a zero-temperature distribution within the bands. This assumption is mainly for convenience, while our conclusions do not depend sensitively on the effective temperature.
The resulting sliding energy landscapes are shown in Fig.~\ref{fn}, where we compare off-equilibrium photodoping with equilibrium hole doping (dashed lines). Note that photodoping excites $2f$ electrons per unit cell into the conduction bands, while creating the same density of valence holes. We therefore compare this to a doping of $4f$ holes per unit cell. Although chemical doping flattens the energy landscape, too, only photodoping beyond $f=0.5$ actually destabilizes the AA stacking. 
Interestingly, the local minimum at AB stacking remains robust up to very strong photodoping. This is ensured by the remaining bonding bands, while the density of states of the nearly metallic phase near the chemical potential is not very sensitive to sliding. 
Within a substantial window of photodoping levels the AA minimum is thus destabilized and the system slides spontaneously towards the stable AB minimum. 

\begin{figure}
\includegraphics[scale=0.77]{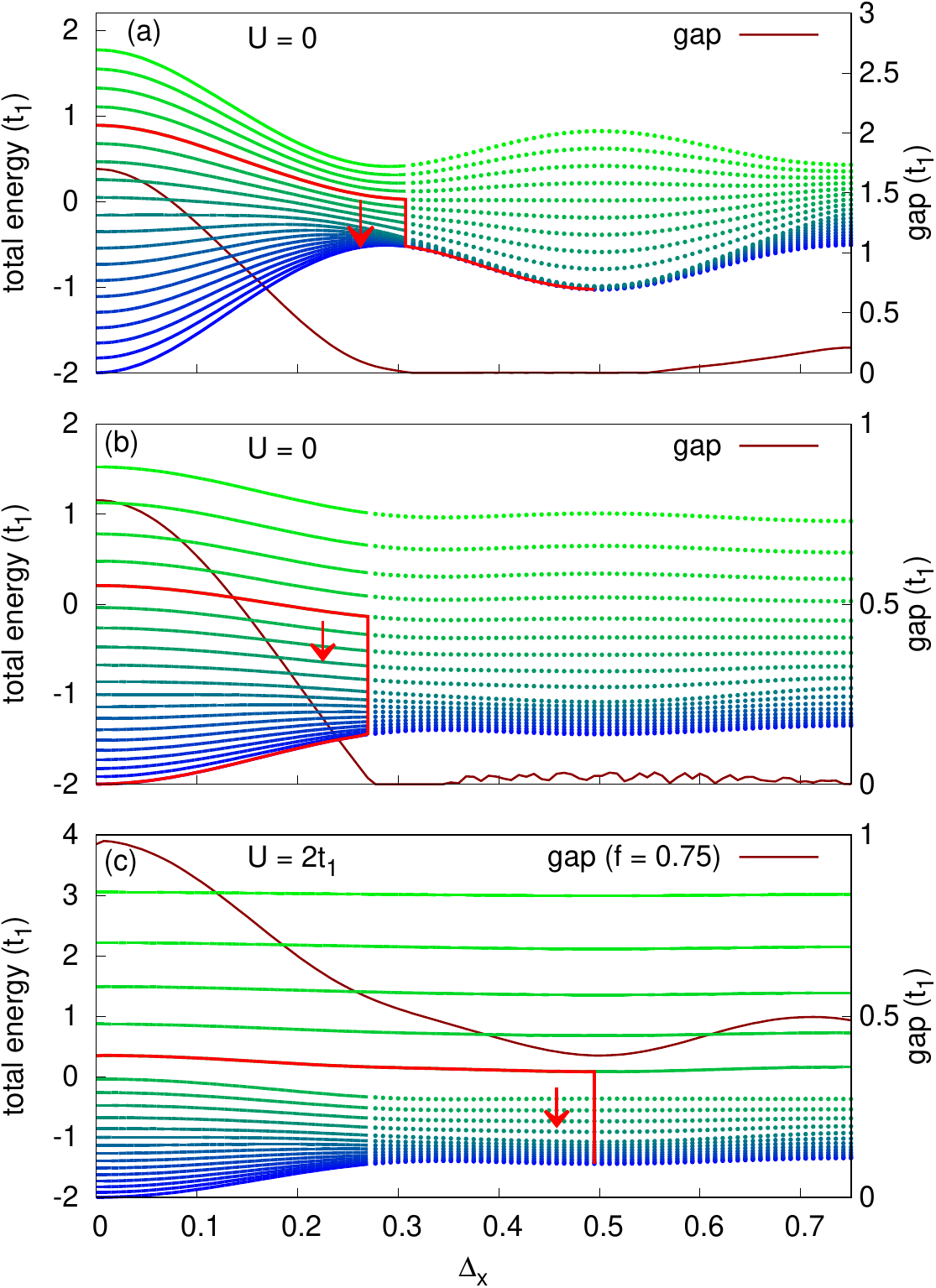}
\caption{Non-equilibrium sliding into a minimum for different parameters. Energies are given in units of $t_1$. (a) $t_1=20t_0$, $\xi=0.33b$, $U=0$. (b) $t_1=10t_0$ and $\xi=b$, $U=0$. (c) Same as (b), but with $U=2t_1$. We use $g_* = 0.2 {t_0}$ as the minimal gap preventing fast recombination.  The initial excited fraction $f$ increases by steps of  0.05 from 0 to 0.95 (blue to green). The solid part of a curve corresponds to states with stable inversion (gap $> g_*$), while photodoping relaxes quickly in the dotted regions. Interactions delay the onset of fast recombination.}
\label{f}
\end{figure}

\emph{Sliding to the metastable phase --} 
To analyze the ensuing dynamics of the photo-excited state, 
we assume the following separation of timescales: As long as a large gap suppresses recombination, the relaxation time of $f$, $\tau_{\text{reco}}$, is much longer than the timescale for sliding $\tau_{\text{slid}}$, while both exceed the intraband thermalization timescale $\tau_{\text{ther}}$ in the nearly metallic phase:
$\tau_{\text{reco}}\gg \tau_{\text{slid}} \gg \tau_{\text{ther}}$. With the large gap near AA stacking, the layers will then slide at essentially fixed $f$.  Upon sliding the gap decreases, and we assume that once it drops below a threshold $g_*$, 
recombination becomes fast
and $f$ decreases rapidly (on the timescale $\tau_\text{ther}$).
The value of $g_*$ depends on the relevant recombination mechanism. If recombination is induced by phonon scattering, $g_*$ will be of the order of the Debye frequency $\omega_D$.

These considerations suggest the following heuristic rules for the stacking dynamics: (i) The bilayer slides from the destabilized AA stacking toward the non-equilibrium minimum until the gap drops below  
{$g_*$}.
(ii) While the gap remains large, $f$ hardly decreases, except if (iii) the system settles into a nonthermal, insulating minimum. In that case $f$ slowly relaxes, while the shift vector adiabatically follows the minimum, potentially until $f$ relaxes to 0. (iv) Once the gap falls below 
{$g_*$}, 
recombination is enhanced and $f$ rapidly drops to 0. Afterwards, $\Delta$ follows the gradient of the equilibrium energy landscape to the nearest local minimum, whereby we assume the electron temperature to remain close to the bath temperature, which we set to $T\approx 0$. 

A typical trajectory determined by the above rules is shown in Fig.~\ref{f}(a). Regions with a large gap are plotted with solid lines while dots indicate a gap $<g_*$. After photodoping a fraction $f=0.75$ of electrons into the valence band, the layered system slides from the destabilized AA toward the stable AB stacking. As the gap falls below the threshold $g_*$ at some shift $\Delta^*$,
 the electrons quickly recombine, relaxing to $f=0$. If $\Delta^*$ belongs to the basin of attraction of the AB-stacked metastable minimum the system slides into it. However, depending on parameters, $\Delta^*$ might still be too small, so that the layers slide back to AA stacking, as shown in Fig.~\ref{f}(b). 

\begin{figure}
\includegraphics[scale=0.5]{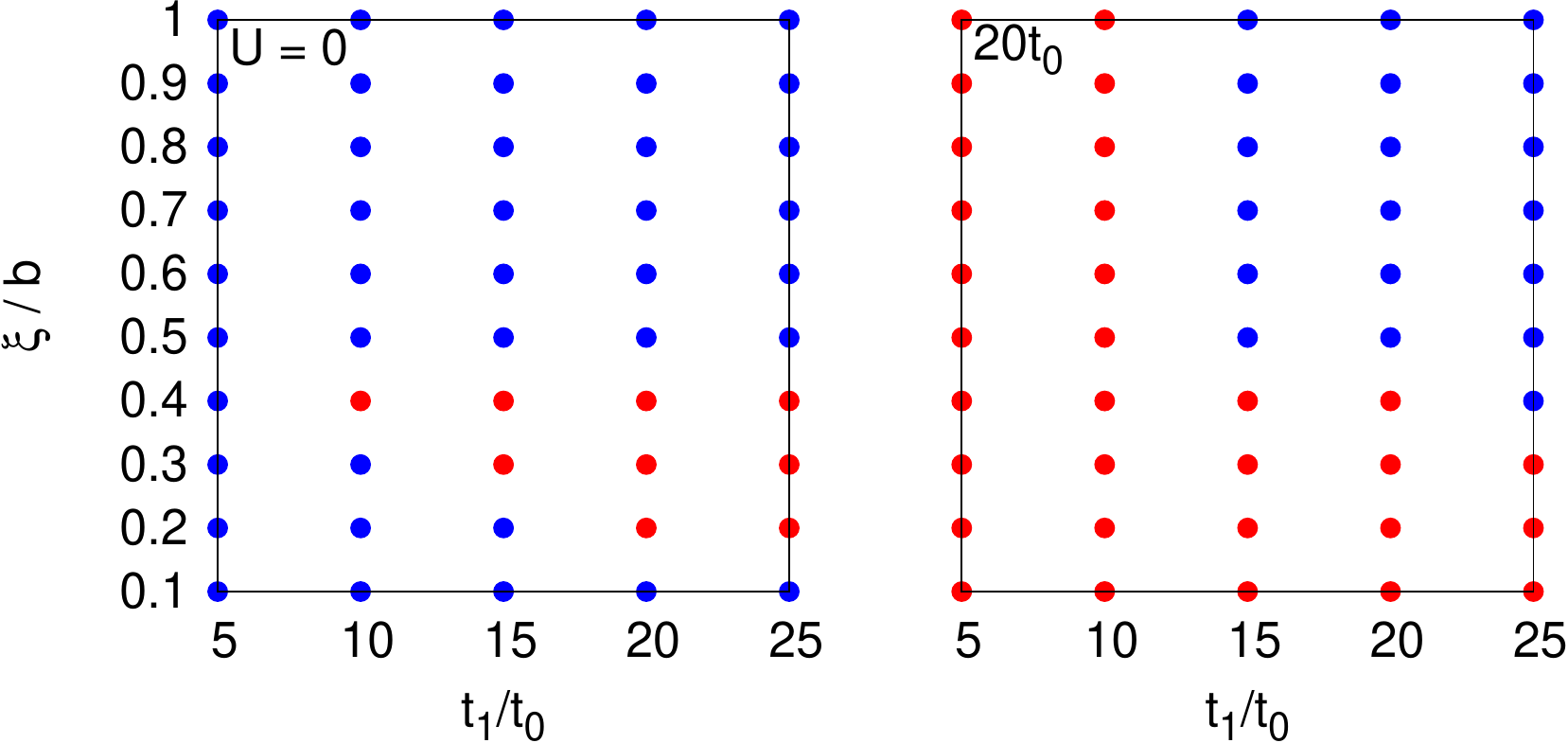}
\caption{Switching diagram of photo-induced sliding for a fraction  $f=0.75$ of photodoped electrons and a {gap} threshold $g_*= 0.2t_0$. Red dots indicate that the hidden phase (AB stacking) is induced deterministically, whereas the system slides back to the AA stacked ground state in the blue region. Left: non-interacting case. Right: a local repulsion $U= 20 t_0$ significantly enlarges the successfully switching region. 
}
\label{pd}
\end{figure}

The hopping parameters for which photo-induced switching to the hidden state succeeds are shown in Fig.~\ref{pd}(a), where we assume an initial photo-doped fraction of $f=0.75$ to destabilize the global minimum, and a small value of the threshold, $g_*= 0.2t_0$. Without interactions, 
a relatively short decay length of the hoppings, $\xi/b\lesssim 0.4 $, and fairly strong interlayer hoppings $t_1/t_0$ are required to ensure a pronounced shift-dependent difference between bonding and anti-bonding states and to prevent recombination to set in before the basin of attraction for AB-stacking (at $f=0$) is reached. 
The situation changes significantly upon turning on a moderate repulsion $U=20t_0$. Computing the sliding energy landscape at different $f$'s using the Hartree-Fock approximation (cf. Fig.~\ref{pd}(b)), we find a substantial interaction gap for large $f$. As it exceeds $g_*$, it stabilizes the population inversion to larger $\Delta$ and thus allows the system to reach the AB stacked local minimum before fast recombination sets in.
This substantially extends the regime of successful  switching to the hidden phase (red points).
The interaction gap originates from the strong charge imbalance between the two atoms in the unit cell, which become inequivalent under sliding. The interactions also increase the barrier separating the secondary from the global minimum.
This arises because a deviation from AB stacking weakens the hopping, which translates into an enhanced interaction gap and thus a higher barrier, as can be seen by comparing the energy landscapes for finite and vanishing $U$. 
While the range of deterministically switching non-interacting systems shrinks with increasing threshold $g_*$  (see SM), 
 the stabilization due to interactions is robust and independent of small $g_*$. 

\emph{Conclusion and Discussion --} Using a minimal electronic model, we have demonstrated that metastable phases exist ubiquitously in the kinetic energy landscape of stacked bilayers with more than one atom per unit cell, and that the stability of stackings can be reversed by a significant non-equilibrium redistribution of electrons between occupied and empty bands using photodoping. In contrast to chemical doping which can only flatten the energy landscape, photodoping is able to destabilize the global minimum upon inverting the band population, inducing spontaneous sliding into a metastable stacking configuration. This provides a pathway for ultrafast optically controlled switching to hidden phases of layered systems.
Moderate repulsions further enhance the robustness of the switching by creating a nonequilibrium interaction gap that suppresses fast recombination and preserves the electronic population inversion while the layers slide toward their metastable stacking configuration. However, since stronger interactions entail faster Auger recombination, the relevant value of $g_*$ may increase with $U$, with the danger of fast recombination setting in too early. In such cases, one may have to resort to continuous photo-doping to nevertheless maintain the population inversion during the sliding process.

A uniform sliding of macroscopic layers is unlikely. Indeed, since there are several equivalent sliding directions and secondary minima, 
different domains may slide in different directions, resulting in a mosaic-like structure \cite{ma2016}. 

The present study was partially motivated by the hidden phase of 1$T$-TaS$_2$. Despite the simplicity of our model, some of our findings are qualitatively consistent with experimental observations: For example, it has been shown that lattice defects and doping with Ti help to form the metallic~\footnote{Note  that our model contains an even number of electrons per unit cell while for 1T-TaS$_2$ it is odd. This leads to a metallic (rather than semiconducting) secondary minimum. This difference should, however, not change the  sliding scenario qualitatively.} hidden state \cite{salzmann2022,zhang2022}, which can be rationalized by the flattening of the energy surface by chemical doping. The laser driving employed so far, for instance in Ref.~\citenum{stojchevska2014}, may not have been able to create a strongly inverted population, and the resulting weak photodoping has probably had an effect resembling that of  chemical doping.
Our results, however, predict that an inverted population, created by strong and carefully designed pulses with a chirped frequency that tracks the difference between the two effective chemical potentials \cite{werner2019}, can facilitate the formation of the hidden phase, possibly on yet faster time scales than other excitation protocols. As strong inverted band populations can overcome even high energy barriers, our findings also suggest an alternative  non-thermal pathway to induce sliding interfacial ferroelectricity in a wider range of materials.

The present theoretical description could be extended 
to incorporate {\it ab initio} input, replacing atomic sites with Wannier orbitals from first-principles calculations.
While we expect our heuristic rules to  predict the sliding pathway correctly,  
a quantitative prediction of the switching time would require time-dependent simulations for the combined lattice and electronic dynamics. This could be achieved with Quantum Boltzmann equation methods assisted by dynamical mean-field theory \cite{picano2021prb,picano2021} or Ehrenfest dynamics within time-dependent density functional theory \cite{tancogne2020}.

The sliding of layered materials can be seen as a solid-state analogue of photoisomerization, i.e.,  molecular conformation changes induced by electronic excitations  \cite{garavelli1997}. 
Those often owe their efficiency to conical intersections of the ground and excited energy surfaces, which funnel the dynamics of the atoms toward a `hidden' conformation~\cite{Martinez2010}. This contrasts with our case where the system slides to a secondary minimum without any intersection, the closing of the band gap being preferentially prevented rather than promoted.  
However, it would be interesting to explore whether the mechanism we propose for solids, and in particular its boost by interactions, have analogues in molecules or nanophysics. 

\begin{acknowledgments}
We acknowledge helpful discussions with D. Baeriswyl, F. Petocchi, M. Eckstein, J. Maklar, and L. Rettig. The project was supported by funding from the European Union’s Horizon 2020 research and innovation programme under the Marie Sklodowska-Curie grant
agreement No. 884104, from the ERC Consolidator Grant No. 724103 and from Swiss National Science Foundation Grants Nos. 200021-196966 and 200020$\_$200558. MAS acknowledges financial support through the Deutsche Forschungsgemeinschaft (DFG, German Research Foundation) via the Emmy Noether program (SE 2558/2).
\end{acknowledgments}

\bibliography{ref.bib}

\onecolumngrid
\appendix
\clearpage

\begin{center}
\textbf{Supplementary Materials}
\end{center}
\section{Details of the model}
The hopping amplitudes are assumed to be simple functions of the atomic distance, $t_{i j;\alpha\alpha'}=t_{\alpha\alpha'}\exp(-(d_{ij;\alpha\alpha'}-d_{{\rm min};\alpha\alpha'})/\xi)$. The shortest distances between atoms from the same layer and from different layers are $d_{{\rm min};11}=d_{{\rm min};22}$ and $d_{{\rm min};12}$, respectively \footnote{Usually there are additional directional factors taking into account the orientation of chemical bonds. We neglect such factors for simplicity. }.
We take $\bm a_1,\bm a_2$ to be two basis vectors of the underlying Bravais lattice. The location of a site $i$ is then $\bm R_{i} = n_{i1}\bm a_1 + n_{i2}\bm a_2 +\bm \delta_i$, where $\bm \delta_{i}$ is the relative location of site $i$ in the unit cell.

For intralayer distances,
we always have
\begin{align}
d_{ij;\alpha=\alpha'}&=\sqrt{((\bm n_{i}-\bm n_{j}+\bm\delta_{ij})\cdot \hat{\bm x})^2+((\bm n_{i}-\bm n_{j}+\bm\delta_{ij})\cdot \hat{\bm y})^2}\nonumber\\
&=[((n_{i1}-n_{j1}+\delta_{ij,1})a_{1x}+(n_{i2}-n_{j2}+\delta_{ij,2})a_{2x})^2+((n_{i1}-n_{j1}+\delta_{ij,1})a_{1y}+(n_{i2}-n_{j2}+\delta_{ij,2})a_{2y})^2]^{1/2},
\end{align}
where $\bm \delta_{ij} = \bm \delta_{i}-\bm \delta_j$. Assuming a relative shift $\bm \Delta$ between the layers,
the interlayer distance
\begin{align}
d_{ij;12}&=\sqrt{((\bm n_i-\bm n_j-\bm \Delta+\bm\delta_{ij})\cdot \hat{\bm x})^2+((\bm n_i-\bm n_j-\bm \Delta+\bm\delta_{ij})\cdot \hat{\bm y})^2+b^2}\nonumber\\
&=[((n_{i1}-n_{j1}-\Delta_1+\delta_{ij,1})a_{1x}+(n_{i2}-n_{j2}-\Delta_2+\delta_{ij,1})a_{2x})^2+\nonumber\\
&((n_{i1}-n_{j1}-\Delta_1+\delta_{ij,1})a_{1y}+(n_{i2}-n_{j2}-\Delta_2+\delta_{ij,2})a_{2y})^2+b^2]^{1/2}.
\end{align}
In general, the interlayer spacing $b$ might itself change slightly with $\bm \Delta$, but we neglect such a  dependence for simplicity. 

\section{Robustness of the secondary minimum}
\begin{figure*}[h]
\includegraphics[scale=2]{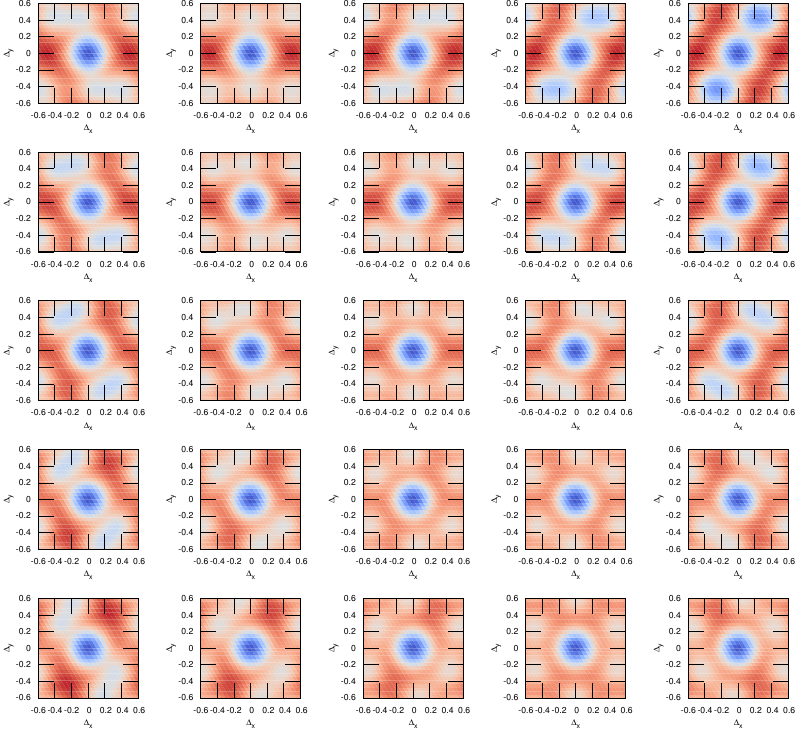}
\caption{The energy landscape of a   bilayer of {distorted} hexagonal lattices, where the B atom takes  different positions $(\alpha_B,\beta_B)$ in the unit cell. The columns correspond to $\alpha_B=0.20,0.25,\ldots,0.40$, while the rows correspond to $\beta_B=0.50,0.55,\ldots,0.70$. The hopping parameters are taken to be $\xi=0.4b$ and $t_1=40t_0$.}
\label{exy}
\end{figure*}

In the main text we have focused on {an undistorted honeycomb lattice. However, the existence of the local minima is robust with respect to distortions of the lattice.} In Fig.~\ref{exy} we show the energy landscape to exhibit a secondary minimum for a wide set of different positions of the B atom within the unit cell. Meanwhile, the sublattice of the A atoms is fixed to remain triangular.  

\section{Inverted energy landscape under negative temperatures}

Here we show that a negative-temperature distribution 
 always reverses the sign of the slope of the energy landscape and thus destabilizes all minima. In  the limit of infinite negative temperature $\beta\to \infty$ this describes the case of a full band inversion due to extreme photodoping ($f\to 1$), and thus shows that the latter destabilizes all minima.
 
 We assume $H=H_0$ and $t_{ij;\alpha\bar{\alpha}}=t_{ji;\bar{\alpha}\alpha}$. The derivative of the energy w.r.t $\bm \Delta$ is
\begin{align}
\partial \langle H_0\rangle/\partial \bm \Delta=-\sum_{i,j,\alpha,\bar{\alpha},\sigma}(\partial t_{ij;\alpha\bar{\alpha}}/\partial\bm\Delta) \langle c^\dag_{i\alpha\sigma}c_{j\bar{\alpha}\sigma}\rangle.
\end{align}
Using the subscript $\beta$ to denote the temperature under which the expectation is taken, we note that
\begin{align}
\langle c^\dag_{i\alpha\sigma}c_{j\bar{\alpha}\sigma} \rangle_\beta=\operatorname{Tr}(e^{-\beta(H_0-\mu N)}c^\dag_{i\alpha\sigma}c_{j\bar{\alpha}\sigma}),
\end{align}
where $N=\sum_{i\alpha\sigma}c^\dag_{i\alpha\sigma}c_{i\alpha\sigma}$. 

We consider the system at fixed chemical potential $\mu$.
Applying a particle-hole transformation, $P^\dag c^\dag_{i\alpha\sigma} P=c_{i\alpha\sigma}$, leads to $P^\dag H_0P=-H_0$ and $P^\dag NP=2n_{\rm lat} - N$, where $n_{\rm lat}$ is the number of lattice sites. 
We then have , 
\begin{align}
\langle c^\dag_{i\alpha\sigma}c_{j\bar{\alpha}\sigma} \rangle_\beta&=\operatorname{Tr}(P^\dag e^{-\beta(H_0-\mu N)}c^\dag_{i\alpha\sigma}c_{j\bar{\alpha}\sigma}P)\nonumber\\
&=\operatorname{Tr}( e^{-\beta(-H_0-2\mu n_{\rm lat}+\mu N)}c_{i\alpha\sigma}c^\dag_{j\bar{\alpha}\sigma})\nonumber\\
&=-e^{2\mu\beta n_{\rm lat}}\operatorname{Tr}( e^{\beta(H_0-\mu N)}c^\dag_{j\bar{\alpha}\sigma}c_{i\alpha\sigma})\nonumber\\
&=-e^{2\mu\beta n_{\rm lat}}\langle c^\dag_{j\bar{\alpha}\sigma}c_{i\alpha\sigma} \rangle_{-\beta}.
\end{align}

From the above equality we obtain 
\begin{align}
\left.\partial \langle H_0\rangle/\partial \bm \Delta\right|_{\beta} = \sum_{ij \alpha\sigma}\frac{\partial t_{ij;\alpha\bar{\alpha}}}{\partial\bm \Delta} \langle c^\dag_{i\alpha\sigma}c_{j\bar{\alpha}\sigma} \rangle_\beta&=\frac{1}{2}\sum_{ij \alpha\sigma}\frac{\partial t_{ij;\alpha\bar{\alpha}}}{\partial\bm \Delta} (\langle c^\dag_{i\alpha\sigma}c_{j\bar{\alpha}\sigma} \rangle_\beta+\langle c^\dag_{j\bar{\alpha}\sigma}c_{i\alpha\sigma} \rangle_\beta)\nonumber\\
&=-e^{2\mu\beta n_{\rm lat}}\frac{1}{2}\sum_{ij \alpha\sigma}\frac{\partial t_{ij;\alpha\bar{\alpha}}}{\partial\bm \Delta} (\langle c^\dag_{j\bar{\alpha}\sigma}c_{i\alpha\sigma} \rangle_{-\beta}+\langle c^\dag_{i\alpha\sigma}c_{j\bar{\alpha}\sigma} \rangle_{-\beta})\nonumber\\
&=-e^{2\mu\beta n_{\rm lat}}\sum_{ij \alpha\sigma}\frac{\partial t_{ij;\alpha\bar{\alpha}}}{\partial\bm \Delta} \langle c^\dag_{i\alpha\sigma}c_{j\bar{\alpha}\sigma} \rangle_{-\beta},\nonumber\\
&= -e^{2\mu\beta n_{\rm lat}} 
\left.\partial \langle H_0\rangle/\partial \bm \Delta\right|_{-\beta}.
\end{align}
This shows that the slope of the 
energy landscape in any direction has a sign opposite to that corresponding to positive temperature.

\section{Supplemental data on the dependence of the switching diagram on the gap threshold $g_*$}

Here we study how, in the non-interacting case, the  switching diagram depends on the value of the phonon threshold $g_*$. 

\begin{figure}
    \centering
        \includegraphics{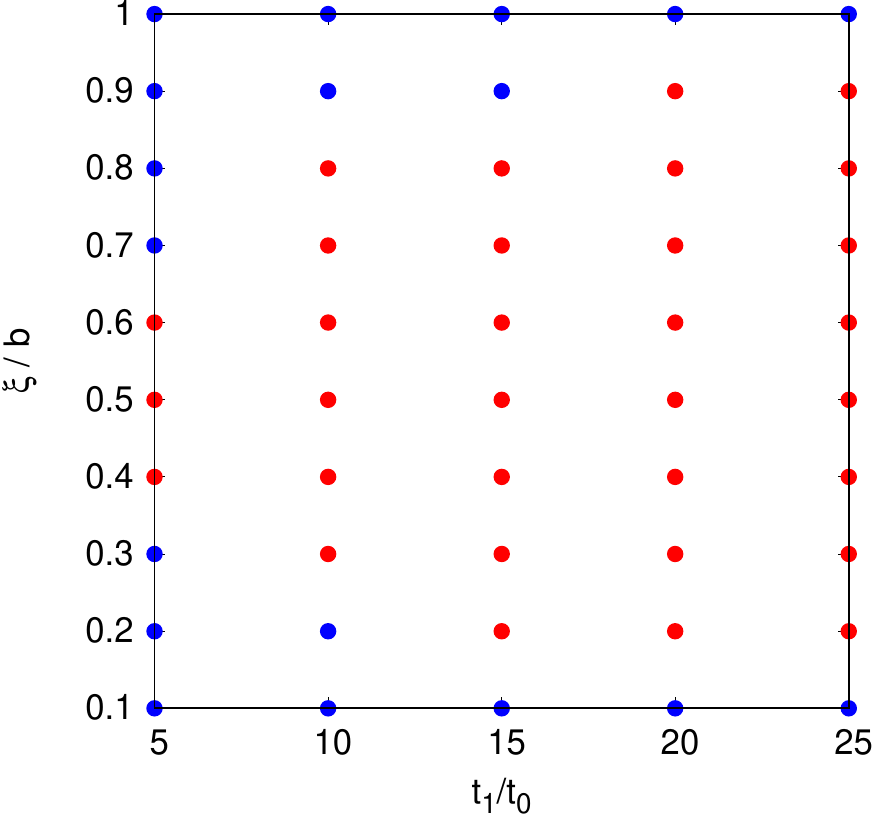}\includegraphics{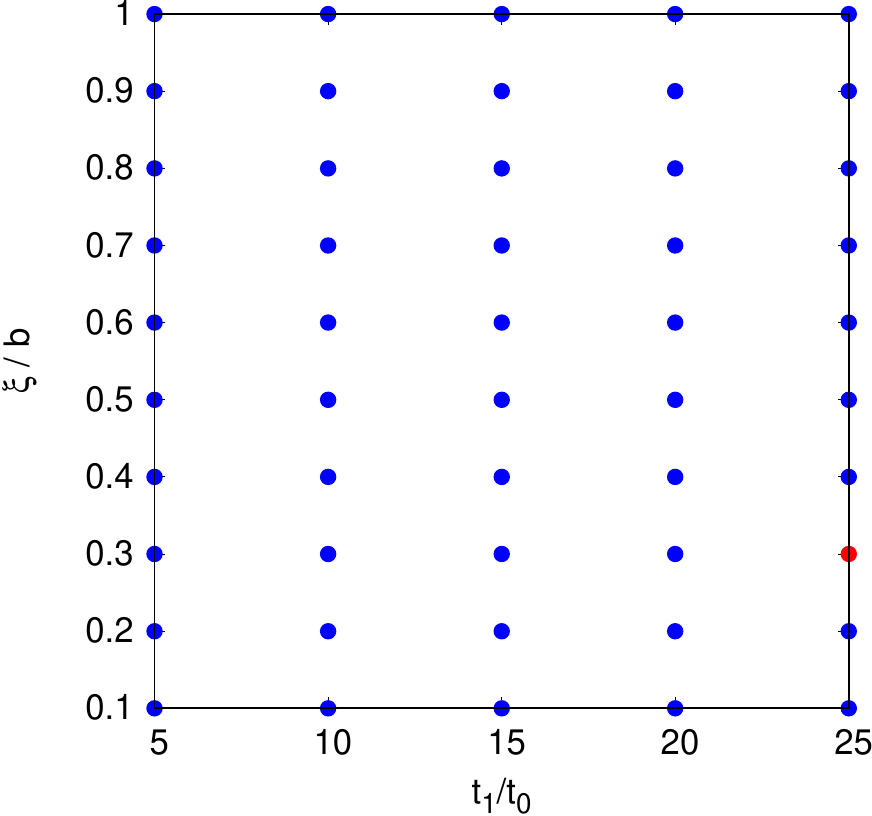}
    \caption{The switching phase diagram in the non-interacting case ($U=0$). {\it Left:} Recombination is assumed to never speed up, i.e., the critical gap is assumed to be $g_*=0$ (in the numerics we used the value $g_* = 10^{-9}t_0$). {\it Right:} Recombination is assumed to speed up when the gap falls below the value $g_*=t_0$. We considered the same lattice as  in Fig.~\ref{pd}, and the same initial photodoping level  of $f=0.75$}
    \label{pd3}
\end{figure}

On the left of Fig.~\ref{pd3} we show data obtained for the case where we assume that recombination never becomes faster than the sliding motion, corresponding to a threshold gap $g_*=0$, which is never reached. 
In this case the switching succeeds in the largest part of parameter space. It fails only if the system does not slide all the way to the minimum that turns into the hidden state upon adiabatic decrease of $f$, but instead sticks  to another minimum which eventually evolves back to the ground state as $f\to 0$.

We then analyze the case of a gap threshold $g_* = t_0$, which is substantially larger than the value $g_* = 0.2 t_0$ considered in Fig.~\ref{pd}
In this case only a tiny pocket of hopping parameter space with very strong bonding tendency still succeeds in switching deterministically.

In this case the deterministic switching of the non-interacting system fails due to the premature de-excitation once the gap reaches the threshold $g_*$. However, upon de-excitation the system falls close to the maximum of the ground state energy landscape ($f=0$). Hence, if the de-excitation is not extremely fast compared to sliding time scales, or if fluctuation effects play a role, there may still be a chance that the system slides into the hidden phase, though perhaps just with a finite probability instead of with near certainty. This situation resembles what happens in photoisomerization processes in molecules that are funneled through conical intersections. Also there the evolution is not fully deterministic, but the conformation change has a finite failure rate.

\bibliography{ref.bib}

\begin{thebibliography}{40}%
\makeatletter
\providecommand \@ifxundefined [1]{%
 \@ifx{#1\undefined}
}%
\providecommand \@ifnum [1]{%
 \ifnum #1\expandafter \@firstoftwo
 \else \expandafter \@secondoftwo
 \fi
}%
\providecommand \@ifx [1]{%
 \ifx #1\expandafter \@firstoftwo
 \else \expandafter \@secondoftwo
 \fi
}%
\providecommand \natexlab [1]{#1}%
\providecommand \enquote  [1]{``#1''}%
\providecommand \bibnamefont  [1]{#1}%
\providecommand \bibfnamefont [1]{#1}%
\providecommand \citenamefont [1]{#1}%
\providecommand \href@noop [0]{\@secondoftwo}%
\providecommand \href [0]{\begingroup \@sanitize@url \@href}%
\providecommand \@href[1]{\@@startlink{#1}\@@href}%
\providecommand \@@href[1]{\endgroup#1\@@endlink}%
\providecommand \@sanitize@url [0]{\catcode `\\12\catcode `\$12\catcode
  `\&12\catcode `\#12\catcode `\^12\catcode `\_12\catcode `\%12\relax}%
\providecommand \@@startlink[1]{}%
\providecommand \@@endlink[0]{}%
\providecommand \url  [0]{\begingroup\@sanitize@url \@url }%
\providecommand \@url [1]{\endgroup\@href {#1}{\urlprefix }}%
\providecommand \urlprefix  [0]{URL }%
\providecommand \Eprint [0]{\href }%
\providecommand \doibase [0]{https://doi.org/}%
\providecommand \selectlanguage [0]{\@gobble}%
\providecommand \bibinfo  [0]{\@secondoftwo}%
\providecommand \bibfield  [0]{\@secondoftwo}%
\providecommand \translation [1]{[#1]}%
\providecommand \BibitemOpen [0]{}%
\providecommand \bibitemStop [0]{}%
\providecommand \bibitemNoStop [0]{.\EOS\space}%
\providecommand \EOS [0]{\spacefactor3000\relax}%
\providecommand \BibitemShut  [1]{\csname bibitem#1\endcsname}%
\let\auto@bib@innerbib\@empty
\bibitem [{\citenamefont {Basov}\ \emph {et~al.}(2017)\citenamefont {Basov},
  \citenamefont {Averitt},\ and\ \citenamefont {Hsieh}}]{basov2017}%
  \BibitemOpen
  \bibfield  {author} {\bibinfo {author} {\bibfnamefont {D.}~\bibnamefont
  {Basov}}, \bibinfo {author} {\bibfnamefont {R.}~\bibnamefont {Averitt}},\
  and\ \bibinfo {author} {\bibfnamefont {D.}~\bibnamefont {Hsieh}},\ }\bibfield
   {title} {\bibinfo {title} {Towards properties on demand in quantum
  materials},\ }\href@noop {} {\bibfield  {journal} {\bibinfo  {journal}
  {Nature materials}\ }\textbf {\bibinfo {volume} {16}},\ \bibinfo {pages}
  {1077} (\bibinfo {year} {2017})}\BibitemShut {NoStop}%
\bibitem [{\citenamefont {de~la Torre}\ \emph {et~al.}(2021)\citenamefont
  {de~la Torre}, \citenamefont {Kennes}, \citenamefont {Claassen},
  \citenamefont {Gerber}, \citenamefont {McIver},\ and\ \citenamefont
  {Sentef}}]{delatorre2021}%
  \BibitemOpen
  \bibfield  {author} {\bibinfo {author} {\bibfnamefont {A.}~\bibnamefont
  {de~la Torre}}, \bibinfo {author} {\bibfnamefont {D.~M.}\ \bibnamefont
  {Kennes}}, \bibinfo {author} {\bibfnamefont {M.}~\bibnamefont {Claassen}},
  \bibinfo {author} {\bibfnamefont {S.}~\bibnamefont {Gerber}}, \bibinfo
  {author} {\bibfnamefont {J.~W.}\ \bibnamefont {McIver}},\ and\ \bibinfo
  {author} {\bibfnamefont {M.~A.}\ \bibnamefont {Sentef}},\ }\bibfield  {title}
  {\bibinfo {title} {Colloquium: Nonthermal pathways to ultrafast control in
  quantum materials},\ }\href {https://doi.org/10.1103/RevModPhys.93.041002}
  {\bibfield  {journal} {\bibinfo  {journal} {Rev. Mod. Phys.}\ }\textbf
  {\bibinfo {volume} {93}},\ \bibinfo {pages} {041002} (\bibinfo {year}
  {2021})}\BibitemShut {NoStop}%
\bibitem [{\citenamefont {Geim}\ and\ \citenamefont
  {Grigorieva}(2013)}]{geim2013}%
  \BibitemOpen
  \bibfield  {author} {\bibinfo {author} {\bibfnamefont {A.~K.}\ \bibnamefont
  {Geim}}\ and\ \bibinfo {author} {\bibfnamefont {I.~V.}\ \bibnamefont
  {Grigorieva}},\ }\bibfield  {title} {\bibinfo {title} {Van der waals
  heterostructures},\ }\href@noop {} {\bibfield  {journal} {\bibinfo  {journal}
  {Nature}\ }\textbf {\bibinfo {volume} {499}},\ \bibinfo {pages} {419}
  (\bibinfo {year} {2013})}\BibitemShut {NoStop}%
\bibitem [{\citenamefont {Novoselov}\ \emph {et~al.}(2016)\citenamefont
  {Novoselov}, \citenamefont {Mishchenko}, \citenamefont {Carvalho},\ and\
  \citenamefont {Castro~Neto}}]{novoselov2016}%
  \BibitemOpen
  \bibfield  {author} {\bibinfo {author} {\bibfnamefont {K.}~\bibnamefont
  {Novoselov}}, \bibinfo {author} {\bibfnamefont {o.~A.}\ \bibnamefont
  {Mishchenko}}, \bibinfo {author} {\bibfnamefont {o.~A.}\ \bibnamefont
  {Carvalho}},\ and\ \bibinfo {author} {\bibfnamefont {A.}~\bibnamefont
  {Castro~Neto}},\ }\bibfield  {title} {\bibinfo {title} {2d materials and van
  der waals heterostructures},\ }\href@noop {} {\bibfield  {journal} {\bibinfo
  {journal} {Science}\ }\textbf {\bibinfo {volume} {353}},\ \bibinfo {pages}
  {aac9439} (\bibinfo {year} {2016})}\BibitemShut {NoStop}%
\bibitem [{\citenamefont {Cao}\ \emph {et~al.}(2018{\natexlab{a}})\citenamefont
  {Cao}, \citenamefont {Fatemi}, \citenamefont {Fang}, \citenamefont
  {Watanabe}, \citenamefont {Taniguchi}, \citenamefont {Kaxiras},\ and\
  \citenamefont {Jarillo-Herrero}}]{cao2018}%
  \BibitemOpen
  \bibfield  {author} {\bibinfo {author} {\bibfnamefont {Y.}~\bibnamefont
  {Cao}}, \bibinfo {author} {\bibfnamefont {V.}~\bibnamefont {Fatemi}},
  \bibinfo {author} {\bibfnamefont {S.}~\bibnamefont {Fang}}, \bibinfo {author}
  {\bibfnamefont {K.}~\bibnamefont {Watanabe}}, \bibinfo {author}
  {\bibfnamefont {T.}~\bibnamefont {Taniguchi}}, \bibinfo {author}
  {\bibfnamefont {E.}~\bibnamefont {Kaxiras}},\ and\ \bibinfo {author}
  {\bibfnamefont {P.}~\bibnamefont {Jarillo-Herrero}},\ }\bibfield  {title}
  {\bibinfo {title} {Unconventional superconductivity in magic-angle graphene
  superlattices},\ }\href@noop {} {\bibfield  {journal} {\bibinfo  {journal}
  {Nature}\ }\textbf {\bibinfo {volume} {556}},\ \bibinfo {pages} {43}
  (\bibinfo {year} {2018}{\natexlab{a}})}\BibitemShut {NoStop}%
\bibitem [{\citenamefont {Cao}\ \emph {et~al.}(2018{\natexlab{b}})\citenamefont
  {Cao}, \citenamefont {Fatemi}, \citenamefont {Demir}, \citenamefont {Fang},
  \citenamefont {Tomarken}, \citenamefont {Luo}, \citenamefont
  {Sanchez-Yamagishi}, \citenamefont {Watanabe}, \citenamefont {Taniguchi},
  \citenamefont {Kaxiras} \emph {et~al.}}]{cao2018correl}%
  \BibitemOpen
  \bibfield  {author} {\bibinfo {author} {\bibfnamefont {Y.}~\bibnamefont
  {Cao}}, \bibinfo {author} {\bibfnamefont {V.}~\bibnamefont {Fatemi}},
  \bibinfo {author} {\bibfnamefont {A.}~\bibnamefont {Demir}}, \bibinfo
  {author} {\bibfnamefont {S.}~\bibnamefont {Fang}}, \bibinfo {author}
  {\bibfnamefont {S.~L.}\ \bibnamefont {Tomarken}}, \bibinfo {author}
  {\bibfnamefont {J.~Y.}\ \bibnamefont {Luo}}, \bibinfo {author} {\bibfnamefont
  {J.~D.}\ \bibnamefont {Sanchez-Yamagishi}}, \bibinfo {author} {\bibfnamefont
  {K.}~\bibnamefont {Watanabe}}, \bibinfo {author} {\bibfnamefont
  {T.}~\bibnamefont {Taniguchi}}, \bibinfo {author} {\bibfnamefont
  {E.}~\bibnamefont {Kaxiras}}, \emph {et~al.},\ }\bibfield  {title} {\bibinfo
  {title} {Correlated insulator behaviour at half-filling in magic-angle
  graphene superlattices},\ }\href@noop {} {\bibfield  {journal} {\bibinfo
  {journal} {Nature}\ }\textbf {\bibinfo {volume} {556}},\ \bibinfo {pages}
  {80} (\bibinfo {year} {2018}{\natexlab{b}})}\BibitemShut {NoStop}%
\bibitem [{\citenamefont {Kennes}\ \emph {et~al.}(2021)\citenamefont {Kennes},
  \citenamefont {Claassen}, \citenamefont {Xian}, \citenamefont {Georges},
  \citenamefont {Millis}, \citenamefont {Hone}, \citenamefont {Dean},
  \citenamefont {Basov}, \citenamefont {Pasupathy},\ and\ \citenamefont
  {Rubio}}]{kennes2021}%
  \BibitemOpen
  \bibfield  {author} {\bibinfo {author} {\bibfnamefont {D.~M.}\ \bibnamefont
  {Kennes}}, \bibinfo {author} {\bibfnamefont {M.}~\bibnamefont {Claassen}},
  \bibinfo {author} {\bibfnamefont {L.}~\bibnamefont {Xian}}, \bibinfo {author}
  {\bibfnamefont {A.}~\bibnamefont {Georges}}, \bibinfo {author} {\bibfnamefont
  {A.~J.}\ \bibnamefont {Millis}}, \bibinfo {author} {\bibfnamefont
  {J.}~\bibnamefont {Hone}}, \bibinfo {author} {\bibfnamefont {C.~R.}\
  \bibnamefont {Dean}}, \bibinfo {author} {\bibfnamefont {D.}~\bibnamefont
  {Basov}}, \bibinfo {author} {\bibfnamefont {A.~N.}\ \bibnamefont
  {Pasupathy}},\ and\ \bibinfo {author} {\bibfnamefont {A.}~\bibnamefont
  {Rubio}},\ }\bibfield  {title} {\bibinfo {title} {Moir{\'e} heterostructures
  as a condensed-matter quantum simulator},\ }\href@noop {} {\bibfield
  {journal} {\bibinfo  {journal} {Nature Physics}\ }\textbf {\bibinfo {volume}
  {17}},\ \bibinfo {pages} {155} (\bibinfo {year} {2021})}\BibitemShut
  {NoStop}%
\bibitem [{\citenamefont {Fei}\ \emph {et~al.}(2018)\citenamefont {Fei},
  \citenamefont {Zhao}, \citenamefont {Palomaki}, \citenamefont {Sun},
  \citenamefont {Miller}, \citenamefont {Zhao}, \citenamefont {Yan},
  \citenamefont {Xu},\ and\ \citenamefont {Cobden}}]{fei2018}%
  \BibitemOpen
  \bibfield  {author} {\bibinfo {author} {\bibfnamefont {Z.}~\bibnamefont
  {Fei}}, \bibinfo {author} {\bibfnamefont {W.}~\bibnamefont {Zhao}}, \bibinfo
  {author} {\bibfnamefont {T.~A.}\ \bibnamefont {Palomaki}}, \bibinfo {author}
  {\bibfnamefont {B.}~\bibnamefont {Sun}}, \bibinfo {author} {\bibfnamefont
  {M.~K.}\ \bibnamefont {Miller}}, \bibinfo {author} {\bibfnamefont
  {Z.}~\bibnamefont {Zhao}}, \bibinfo {author} {\bibfnamefont {J.}~\bibnamefont
  {Yan}}, \bibinfo {author} {\bibfnamefont {X.}~\bibnamefont {Xu}},\ and\
  \bibinfo {author} {\bibfnamefont {D.~H.}\ \bibnamefont {Cobden}},\ }\bibfield
   {title} {\bibinfo {title} {Ferroelectric switching of a two-dimensional
  metal},\ }\href@noop {} {\bibfield  {journal} {\bibinfo  {journal} {Nature}\
  }\textbf {\bibinfo {volume} {560}},\ \bibinfo {pages} {336} (\bibinfo {year}
  {2018})}\BibitemShut {NoStop}%
\bibitem [{\citenamefont {Yasuda}\ \emph {et~al.}(2021)\citenamefont {Yasuda},
  \citenamefont {Wang}, \citenamefont {Watanabe}, \citenamefont {Taniguchi},\
  and\ \citenamefont {Jarillo-Herrero}}]{yasuda2021}%
  \BibitemOpen
  \bibfield  {author} {\bibinfo {author} {\bibfnamefont {K.}~\bibnamefont
  {Yasuda}}, \bibinfo {author} {\bibfnamefont {X.}~\bibnamefont {Wang}},
  \bibinfo {author} {\bibfnamefont {K.}~\bibnamefont {Watanabe}}, \bibinfo
  {author} {\bibfnamefont {T.}~\bibnamefont {Taniguchi}},\ and\ \bibinfo
  {author} {\bibfnamefont {P.}~\bibnamefont {Jarillo-Herrero}},\ }\bibfield
  {title} {\bibinfo {title} {Stacking-engineered ferroelectricity in bilayer
  boron nitride},\ }\href@noop {} {\bibfield  {journal} {\bibinfo  {journal}
  {Science}\ }\textbf {\bibinfo {volume} {372}},\ \bibinfo {pages} {1458}
  (\bibinfo {year} {2021})}\BibitemShut {NoStop}%
\bibitem [{\citenamefont {Vizner~Stern}\ \emph {et~al.}(2021)\citenamefont
  {Vizner~Stern}, \citenamefont {Waschitz}, \citenamefont {Cao}, \citenamefont
  {Nevo}, \citenamefont {Watanabe}, \citenamefont {Taniguchi}, \citenamefont
  {Sela}, \citenamefont {Urbakh}, \citenamefont {Hod},\ and\ \citenamefont
  {Ben~Shalom}}]{vizner_stern2021}%
  \BibitemOpen
  \bibfield  {author} {\bibinfo {author} {\bibfnamefont {M.}~\bibnamefont
  {Vizner~Stern}}, \bibinfo {author} {\bibfnamefont {Y.}~\bibnamefont
  {Waschitz}}, \bibinfo {author} {\bibfnamefont {W.}~\bibnamefont {Cao}},
  \bibinfo {author} {\bibfnamefont {I.}~\bibnamefont {Nevo}}, \bibinfo {author}
  {\bibfnamefont {K.}~\bibnamefont {Watanabe}}, \bibinfo {author}
  {\bibfnamefont {T.}~\bibnamefont {Taniguchi}}, \bibinfo {author}
  {\bibfnamefont {E.}~\bibnamefont {Sela}}, \bibinfo {author} {\bibfnamefont
  {M.}~\bibnamefont {Urbakh}}, \bibinfo {author} {\bibfnamefont
  {O.}~\bibnamefont {Hod}},\ and\ \bibinfo {author} {\bibfnamefont
  {M.}~\bibnamefont {Ben~Shalom}},\ }\bibfield  {title} {\bibinfo {title}
  {Interfacial ferroelectricity by van der waals sliding},\ }\href@noop {}
  {\bibfield  {journal} {\bibinfo  {journal} {Science}\ }\textbf {\bibinfo
  {volume} {372}},\ \bibinfo {pages} {1462} (\bibinfo {year}
  {2021})}\BibitemShut {NoStop}%
\bibitem [{\citenamefont {Li}\ and\ \citenamefont {Wu}(2017)}]{li2017}%
  \BibitemOpen
  \bibfield  {author} {\bibinfo {author} {\bibfnamefont {L.}~\bibnamefont
  {Li}}\ and\ \bibinfo {author} {\bibfnamefont {M.}~\bibnamefont {Wu}},\
  }\bibfield  {title} {\bibinfo {title} {Binary compound bilayer and multilayer
  with vertical polarizations: two-dimensional ferroelectrics, multiferroics,
  and nanogenerators},\ }\href@noop {} {\bibfield  {journal} {\bibinfo
  {journal} {ACS nano}\ }\textbf {\bibinfo {volume} {11}},\ \bibinfo {pages}
  {6382} (\bibinfo {year} {2017})}\BibitemShut {NoStop}%
\bibitem [{\citenamefont {Wu}\ and\ \citenamefont {Li}(2021)}]{wu2021}%
  \BibitemOpen
  \bibfield  {author} {\bibinfo {author} {\bibfnamefont {M.}~\bibnamefont
  {Wu}}\ and\ \bibinfo {author} {\bibfnamefont {J.}~\bibnamefont {Li}},\
  }\bibfield  {title} {\bibinfo {title} {Sliding ferroelectricity in 2d van der
  waals materials: Related physics and future opportunities},\ }\href@noop {}
  {\bibfield  {journal} {\bibinfo  {journal} {Proceedings of the National
  Academy of Sciences}\ }\textbf {\bibinfo {volume} {118}},\ \bibinfo {pages}
  {e2115703118} (\bibinfo {year} {2021})}\BibitemShut {NoStop}%
\bibitem [{\citenamefont {Tang}\ and\ \citenamefont {Bauer}(2022)}]{tang2022}%
  \BibitemOpen
  \bibfield  {author} {\bibinfo {author} {\bibfnamefont {P.}~\bibnamefont
  {Tang}}\ and\ \bibinfo {author} {\bibfnamefont {G.~E.}\ \bibnamefont
  {Bauer}},\ }\bibfield  {title} {\bibinfo {title} {The sliding phase
  transition in ferroelectric van der waals bilayers},\ }\href@noop {}
  {\bibfield  {journal} {\bibinfo  {journal} {arXiv preprint arXiv:2208.00442}\
  } (\bibinfo {year} {2022})}\BibitemShut {NoStop}%
\bibitem [{\citenamefont {Nicholson}\ \emph {et~al.}(2022)\citenamefont
  {Nicholson}, \citenamefont {Petocchi}, \citenamefont {Salzmann},
  \citenamefont {Witteveen}, \citenamefont {Rumo}, \citenamefont {Kremer},
  \citenamefont {von Rohr}, \citenamefont {Werner},\ and\ \citenamefont
  {Monney}}]{nicholson2022}%
  \BibitemOpen
  \bibfield  {author} {\bibinfo {author} {\bibfnamefont {C.~W.}\ \bibnamefont
  {Nicholson}}, \bibinfo {author} {\bibfnamefont {F.}~\bibnamefont {Petocchi}},
  \bibinfo {author} {\bibfnamefont {B.}~\bibnamefont {Salzmann}}, \bibinfo
  {author} {\bibfnamefont {C.}~\bibnamefont {Witteveen}}, \bibinfo {author}
  {\bibfnamefont {M.}~\bibnamefont {Rumo}}, \bibinfo {author} {\bibfnamefont
  {G.}~\bibnamefont {Kremer}}, \bibinfo {author} {\bibfnamefont {F.~O.}\
  \bibnamefont {von Rohr}}, \bibinfo {author} {\bibfnamefont {P.}~\bibnamefont
  {Werner}},\ and\ \bibinfo {author} {\bibfnamefont {C.}~\bibnamefont
  {Monney}},\ }\bibfield  {title} {\bibinfo {title} {Modified interlayer
  stacking and insulator to correlated-metal transition driven by uniaxial
  strain in 1$ t $-tas $ \_ $\{$2$\}$ $},\ }\href@noop {} {\bibfield  {journal}
  {\bibinfo  {journal} {arXiv preprint arXiv:2204.05598}\ } (\bibinfo {year}
  {2022})}\BibitemShut {NoStop}%
\bibitem [{\citenamefont {Wu}\ \emph {et~al.}(2022)\citenamefont {Wu},
  \citenamefont {Bu}, \citenamefont {Zhang}, \citenamefont {Fei}, \citenamefont
  {Zheng}, \citenamefont {Gao}, \citenamefont {Luo}, \citenamefont {Liu},
  \citenamefont {Sun},\ and\ \citenamefont {Yin}}]{wu2022}%
  \BibitemOpen
  \bibfield  {author} {\bibinfo {author} {\bibfnamefont {Z.}~\bibnamefont
  {Wu}}, \bibinfo {author} {\bibfnamefont {K.}~\bibnamefont {Bu}}, \bibinfo
  {author} {\bibfnamefont {W.}~\bibnamefont {Zhang}}, \bibinfo {author}
  {\bibfnamefont {Y.}~\bibnamefont {Fei}}, \bibinfo {author} {\bibfnamefont
  {Y.}~\bibnamefont {Zheng}}, \bibinfo {author} {\bibfnamefont
  {J.}~\bibnamefont {Gao}}, \bibinfo {author} {\bibfnamefont {X.}~\bibnamefont
  {Luo}}, \bibinfo {author} {\bibfnamefont {Z.}~\bibnamefont {Liu}}, \bibinfo
  {author} {\bibfnamefont {Y.-P.}\ \bibnamefont {Sun}},\ and\ \bibinfo {author}
  {\bibfnamefont {Y.}~\bibnamefont {Yin}},\ }\bibfield  {title} {\bibinfo
  {title} {Effect of stacking order on the electronic state of
  $1t\text{\ensuremath{-}}{\mathrm{tas}}_{2}$},\ }\href
  {https://doi.org/10.1103/PhysRevB.105.035109} {\bibfield  {journal} {\bibinfo
   {journal} {Phys. Rev. B}\ }\textbf {\bibinfo {volume} {105}},\ \bibinfo
  {pages} {035109} (\bibinfo {year} {2022})}\BibitemShut {NoStop}%
\bibitem [{\citenamefont {Lee}\ \emph {et~al.}(2019)\citenamefont {Lee},
  \citenamefont {Goh},\ and\ \citenamefont {Cho}}]{lee2019}%
  \BibitemOpen
  \bibfield  {author} {\bibinfo {author} {\bibfnamefont {S.-H.}\ \bibnamefont
  {Lee}}, \bibinfo {author} {\bibfnamefont {J.~S.}\ \bibnamefont {Goh}},\ and\
  \bibinfo {author} {\bibfnamefont {D.}~\bibnamefont {Cho}},\ }\bibfield
  {title} {\bibinfo {title} {Origin of the insulating phase and first-order
  metal-insulator transition in $1t\text{\ensuremath{-}}{\mathrm{tas}}_{2}$},\
  }\href {https://doi.org/10.1103/PhysRevLett.122.106404} {\bibfield  {journal}
  {\bibinfo  {journal} {Phys. Rev. Lett.}\ }\textbf {\bibinfo {volume} {122}},\
  \bibinfo {pages} {106404} (\bibinfo {year} {2019})}\BibitemShut {NoStop}%
\bibitem [{\citenamefont {Petocchi}\ \emph {et~al.}(2022)\citenamefont
  {Petocchi}, \citenamefont {Nicholson}, \citenamefont {Salzmann},
  \citenamefont {Pasquier}, \citenamefont {Yazyev}, \citenamefont {Monney},\
  and\ \citenamefont {Werner}}]{petocchi2022}%
  \BibitemOpen
  \bibfield  {author} {\bibinfo {author} {\bibfnamefont {F.}~\bibnamefont
  {Petocchi}}, \bibinfo {author} {\bibfnamefont {C.~W.}\ \bibnamefont
  {Nicholson}}, \bibinfo {author} {\bibfnamefont {B.}~\bibnamefont {Salzmann}},
  \bibinfo {author} {\bibfnamefont {D.}~\bibnamefont {Pasquier}}, \bibinfo
  {author} {\bibfnamefont {O.~V.}\ \bibnamefont {Yazyev}}, \bibinfo {author}
  {\bibfnamefont {C.}~\bibnamefont {Monney}},\ and\ \bibinfo {author}
  {\bibfnamefont {P.}~\bibnamefont {Werner}},\ }\bibfield  {title} {\bibinfo
  {title} {Mott versus hybridization gap in the low-temperature phase of
  $1t\text{\ensuremath{-}}{\mathrm{tas}}_{2}$},\ }\href
  {https://doi.org/10.1103/PhysRevLett.129.016402} {\bibfield  {journal}
  {\bibinfo  {journal} {Phys. Rev. Lett.}\ }\textbf {\bibinfo {volume} {129}},\
  \bibinfo {pages} {016402} (\bibinfo {year} {2022})}\BibitemShut {NoStop}%
\bibitem [{\citenamefont {Stojchevska}\ \emph {et~al.}(2014)\citenamefont
  {Stojchevska}, \citenamefont {Vaskivskyi}, \citenamefont {Mertelj},
  \citenamefont {Kusar}, \citenamefont {Svetin}, \citenamefont {Brazovskii},\
  and\ \citenamefont {Mihailovic}}]{stojchevska2014}%
  \BibitemOpen
  \bibfield  {author} {\bibinfo {author} {\bibfnamefont {L.}~\bibnamefont
  {Stojchevska}}, \bibinfo {author} {\bibfnamefont {I.}~\bibnamefont
  {Vaskivskyi}}, \bibinfo {author} {\bibfnamefont {T.}~\bibnamefont {Mertelj}},
  \bibinfo {author} {\bibfnamefont {P.}~\bibnamefont {Kusar}}, \bibinfo
  {author} {\bibfnamefont {D.}~\bibnamefont {Svetin}}, \bibinfo {author}
  {\bibfnamefont {S.}~\bibnamefont {Brazovskii}},\ and\ \bibinfo {author}
  {\bibfnamefont {D.}~\bibnamefont {Mihailovic}},\ }\bibfield  {title}
  {\bibinfo {title} {Ultrafast switching to a stable hidden quantum state in an
  electronic crystal},\ }\href {https://doi.org/10.1126/science.1241591}
  {\bibfield  {journal} {\bibinfo  {journal} {Science}\ }\textbf {\bibinfo
  {volume} {344}},\ \bibinfo {pages} {177} (\bibinfo {year}
  {2014})}\BibitemShut {NoStop}%
\bibitem [{\citenamefont {Stahl}\ \emph {et~al.}(2020)\citenamefont {Stahl},
  \citenamefont {Kusch}, \citenamefont {Heinsch}, \citenamefont {Garbarino},
  \citenamefont {Kretzschmar}, \citenamefont {Hanff}, \citenamefont
  {Rossnagel}, \citenamefont {Geck},\ and\ \citenamefont
  {Ritschel}}]{stahl2020}%
  \BibitemOpen
  \bibfield  {author} {\bibinfo {author} {\bibfnamefont {Q.}~\bibnamefont
  {Stahl}}, \bibinfo {author} {\bibfnamefont {M.}~\bibnamefont {Kusch}},
  \bibinfo {author} {\bibfnamefont {F.}~\bibnamefont {Heinsch}}, \bibinfo
  {author} {\bibfnamefont {G.}~\bibnamefont {Garbarino}}, \bibinfo {author}
  {\bibfnamefont {N.}~\bibnamefont {Kretzschmar}}, \bibinfo {author}
  {\bibfnamefont {K.}~\bibnamefont {Hanff}}, \bibinfo {author} {\bibfnamefont
  {K.}~\bibnamefont {Rossnagel}}, \bibinfo {author} {\bibfnamefont
  {J.}~\bibnamefont {Geck}},\ and\ \bibinfo {author} {\bibfnamefont
  {T.}~\bibnamefont {Ritschel}},\ }\bibfield  {title} {\bibinfo {title}
  {Collapse of layer dimerization in the photo-induced hidden state of
  1t-tas2},\ }\href@noop {} {\bibfield  {journal} {\bibinfo  {journal} {Nature
  communications}\ }\textbf {\bibinfo {volume} {11}},\ \bibinfo {pages} {1}
  (\bibinfo {year} {2020})}\BibitemShut {NoStop}%
\bibitem [{\citenamefont {Maklar}\ \emph {et~al.}(2022)\citenamefont {Maklar},
  \citenamefont {Dong}, \citenamefont {Sarkar}, \citenamefont {Gerasimenko},
  \citenamefont {Pincelli}, \citenamefont {Beaulieu}, \citenamefont
  {Kirchmann}, \citenamefont {Sobota}, \citenamefont {Yang}, \citenamefont
  {Leuenberger} \emph {et~al.}}]{maklar2022}%
  \BibitemOpen
  \bibfield  {author} {\bibinfo {author} {\bibfnamefont {J.}~\bibnamefont
  {Maklar}}, \bibinfo {author} {\bibfnamefont {S.}~\bibnamefont {Dong}},
  \bibinfo {author} {\bibfnamefont {J.}~\bibnamefont {Sarkar}}, \bibinfo
  {author} {\bibfnamefont {Y.}~\bibnamefont {Gerasimenko}}, \bibinfo {author}
  {\bibfnamefont {T.}~\bibnamefont {Pincelli}}, \bibinfo {author}
  {\bibfnamefont {S.}~\bibnamefont {Beaulieu}}, \bibinfo {author}
  {\bibfnamefont {P.}~\bibnamefont {Kirchmann}}, \bibinfo {author}
  {\bibfnamefont {J.}~\bibnamefont {Sobota}}, \bibinfo {author} {\bibfnamefont
  {S.-L.}\ \bibnamefont {Yang}}, \bibinfo {author} {\bibfnamefont
  {D.}~\bibnamefont {Leuenberger}}, \emph {et~al.},\ }\bibfield  {title}
  {\bibinfo {title} {Coherent light control of a metastable hidden phase},\
  }\href@noop {} {\bibfield  {journal} {\bibinfo  {journal} {arXiv preprint
  arXiv:2206.03788}\ } (\bibinfo {year} {2022})}\BibitemShut {NoStop}%
\bibitem [{\citenamefont {Hollander}\ \emph {et~al.}(2015)\citenamefont
  {Hollander}, \citenamefont {Liu}, \citenamefont {Lu}, \citenamefont {Li},
  \citenamefont {Sun}, \citenamefont {Robinson},\ and\ \citenamefont
  {Datta}}]{hollander2015}%
  \BibitemOpen
  \bibfield  {author} {\bibinfo {author} {\bibfnamefont {M.~J.}\ \bibnamefont
  {Hollander}}, \bibinfo {author} {\bibfnamefont {Y.}~\bibnamefont {Liu}},
  \bibinfo {author} {\bibfnamefont {W.-J.}\ \bibnamefont {Lu}}, \bibinfo
  {author} {\bibfnamefont {L.-J.}\ \bibnamefont {Li}}, \bibinfo {author}
  {\bibfnamefont {Y.-P.}\ \bibnamefont {Sun}}, \bibinfo {author} {\bibfnamefont
  {J.~A.}\ \bibnamefont {Robinson}},\ and\ \bibinfo {author} {\bibfnamefont
  {S.}~\bibnamefont {Datta}},\ }\bibfield  {title} {\bibinfo {title}
  {Electrically driven reversible insulator--metal phase transition in
  1t-tas2},\ }\href@noop {} {\bibfield  {journal} {\bibinfo  {journal} {Nano
  letters}\ }\textbf {\bibinfo {volume} {15}},\ \bibinfo {pages} {1861}
  (\bibinfo {year} {2015})}\BibitemShut {NoStop}%
\bibitem [{\citenamefont {Ma}\ \emph {et~al.}(2016)\citenamefont {Ma},
  \citenamefont {Ye}, \citenamefont {Yu}, \citenamefont {Lu}, \citenamefont
  {Niu}, \citenamefont {Kim}, \citenamefont {Feng}, \citenamefont
  {Tom{\'a}nek}, \citenamefont {Son}, \citenamefont {Chen} \emph
  {et~al.}}]{ma2016}%
  \BibitemOpen
  \bibfield  {author} {\bibinfo {author} {\bibfnamefont {L.}~\bibnamefont
  {Ma}}, \bibinfo {author} {\bibfnamefont {C.}~\bibnamefont {Ye}}, \bibinfo
  {author} {\bibfnamefont {Y.}~\bibnamefont {Yu}}, \bibinfo {author}
  {\bibfnamefont {X.~F.}\ \bibnamefont {Lu}}, \bibinfo {author} {\bibfnamefont
  {X.}~\bibnamefont {Niu}}, \bibinfo {author} {\bibfnamefont {S.}~\bibnamefont
  {Kim}}, \bibinfo {author} {\bibfnamefont {D.}~\bibnamefont {Feng}}, \bibinfo
  {author} {\bibfnamefont {D.}~\bibnamefont {Tom{\'a}nek}}, \bibinfo {author}
  {\bibfnamefont {Y.-W.}\ \bibnamefont {Son}}, \bibinfo {author} {\bibfnamefont
  {X.~H.}\ \bibnamefont {Chen}}, \emph {et~al.},\ }\bibfield  {title} {\bibinfo
  {title} {A metallic mosaic phase and the origin of mott-insulating state in
  1t-tas2},\ }\href@noop {} {\bibfield  {journal} {\bibinfo  {journal} {Nature
  communications}\ }\textbf {\bibinfo {volume} {7}},\ \bibinfo {pages} {1}
  (\bibinfo {year} {2016})}\BibitemShut {NoStop}%
\bibitem [{\citenamefont {Cho}\ \emph {et~al.}(2016)\citenamefont {Cho},
  \citenamefont {Cheon}, \citenamefont {Kim}, \citenamefont {Lee},
  \citenamefont {Cho}, \citenamefont {Cheong},\ and\ \citenamefont
  {Yeom}}]{cho2016}%
  \BibitemOpen
  \bibfield  {author} {\bibinfo {author} {\bibfnamefont {D.}~\bibnamefont
  {Cho}}, \bibinfo {author} {\bibfnamefont {S.}~\bibnamefont {Cheon}}, \bibinfo
  {author} {\bibfnamefont {K.-S.}\ \bibnamefont {Kim}}, \bibinfo {author}
  {\bibfnamefont {S.-H.}\ \bibnamefont {Lee}}, \bibinfo {author} {\bibfnamefont
  {Y.-H.}\ \bibnamefont {Cho}}, \bibinfo {author} {\bibfnamefont {S.-W.}\
  \bibnamefont {Cheong}},\ and\ \bibinfo {author} {\bibfnamefont {H.~W.}\
  \bibnamefont {Yeom}},\ }\bibfield  {title} {\bibinfo {title} {Nanoscale
  manipulation of the mott insulating state coupled to charge order in
  1t-tas2},\ }\href@noop {} {\bibfield  {journal} {\bibinfo  {journal} {Nature
  communications}\ }\textbf {\bibinfo {volume} {7}},\ \bibinfo {pages} {1}
  (\bibinfo {year} {2016})}\BibitemShut {NoStop}%
\bibitem [{\citenamefont {Vaskivskyi}\ \emph {et~al.}(2016)\citenamefont
  {Vaskivskyi}, \citenamefont {Mihailovic}, \citenamefont {Brazovskii},
  \citenamefont {Gospodaric}, \citenamefont {Mertelj}, \citenamefont {Svetin},
  \citenamefont {Sutar},\ and\ \citenamefont {Mihailovic}}]{vaskivskyi2016}%
  \BibitemOpen
  \bibfield  {author} {\bibinfo {author} {\bibfnamefont {I.}~\bibnamefont
  {Vaskivskyi}}, \bibinfo {author} {\bibfnamefont {I.}~\bibnamefont
  {Mihailovic}}, \bibinfo {author} {\bibfnamefont {S.}~\bibnamefont
  {Brazovskii}}, \bibinfo {author} {\bibfnamefont {J.}~\bibnamefont
  {Gospodaric}}, \bibinfo {author} {\bibfnamefont {T.}~\bibnamefont {Mertelj}},
  \bibinfo {author} {\bibfnamefont {D.}~\bibnamefont {Svetin}}, \bibinfo
  {author} {\bibfnamefont {P.}~\bibnamefont {Sutar}},\ and\ \bibinfo {author}
  {\bibfnamefont {D.}~\bibnamefont {Mihailovic}},\ }\bibfield  {title}
  {\bibinfo {title} {Fast electronic resistance switching involving hidden
  charge density wave states},\ }\href@noop {} {\bibfield  {journal} {\bibinfo
  {journal} {Nature communications}\ }\textbf {\bibinfo {volume} {7}},\
  \bibinfo {pages} {1} (\bibinfo {year} {2016})}\BibitemShut {NoStop}%
\bibitem [{\citenamefont {Ravnik}\ \emph {et~al.}(2021)\citenamefont {Ravnik},
  \citenamefont {Diego}, \citenamefont {Gerasimenko}, \citenamefont
  {Vaskivskyi}, \citenamefont {Vaskivskyi}, \citenamefont {Mertelj},
  \citenamefont {Vodeb},\ and\ \citenamefont {Mihailovic}}]{ravnik2021}%
  \BibitemOpen
  \bibfield  {author} {\bibinfo {author} {\bibfnamefont {J.}~\bibnamefont
  {Ravnik}}, \bibinfo {author} {\bibfnamefont {M.}~\bibnamefont {Diego}},
  \bibinfo {author} {\bibfnamefont {Y.}~\bibnamefont {Gerasimenko}}, \bibinfo
  {author} {\bibfnamefont {Y.}~\bibnamefont {Vaskivskyi}}, \bibinfo {author}
  {\bibfnamefont {I.}~\bibnamefont {Vaskivskyi}}, \bibinfo {author}
  {\bibfnamefont {T.}~\bibnamefont {Mertelj}}, \bibinfo {author} {\bibfnamefont
  {J.}~\bibnamefont {Vodeb}},\ and\ \bibinfo {author} {\bibfnamefont
  {D.}~\bibnamefont {Mihailovic}},\ }\bibfield  {title} {\bibinfo {title} {A
  time-domain phase diagram of metastable states in a charge ordered quantum
  material},\ }\href@noop {} {\bibfield  {journal} {\bibinfo  {journal} {Nature
  communications}\ }\textbf {\bibinfo {volume} {12}},\ \bibinfo {pages} {1}
  (\bibinfo {year} {2021})}\BibitemShut {NoStop}%
\bibitem [{\citenamefont {Ligges}\ \emph {et~al.}(2018)\citenamefont {Ligges},
  \citenamefont {Avigo}, \citenamefont {Gole\ifmmode~\check{z}\else
  \v{z}\fi{}}, \citenamefont {Strand}, \citenamefont {Beyazit}, \citenamefont
  {Hanff}, \citenamefont {Diekmann}, \citenamefont {Stojchevska}, \citenamefont
  {Kall\"ane}, \citenamefont {Zhou}, \citenamefont {Rossnagel}, \citenamefont
  {Eckstein}, \citenamefont {Werner},\ and\ \citenamefont
  {Bovensiepen}}]{ligges2018}%
  \BibitemOpen
  \bibfield  {author} {\bibinfo {author} {\bibfnamefont {M.}~\bibnamefont
  {Ligges}}, \bibinfo {author} {\bibfnamefont {I.}~\bibnamefont {Avigo}},
  \bibinfo {author} {\bibfnamefont {D.}~\bibnamefont
  {Gole\ifmmode~\check{z}\else \v{z}\fi{}}}, \bibinfo {author} {\bibfnamefont
  {H.~U.~R.}\ \bibnamefont {Strand}}, \bibinfo {author} {\bibfnamefont
  {Y.}~\bibnamefont {Beyazit}}, \bibinfo {author} {\bibfnamefont
  {K.}~\bibnamefont {Hanff}}, \bibinfo {author} {\bibfnamefont
  {F.}~\bibnamefont {Diekmann}}, \bibinfo {author} {\bibfnamefont
  {L.}~\bibnamefont {Stojchevska}}, \bibinfo {author} {\bibfnamefont
  {M.}~\bibnamefont {Kall\"ane}}, \bibinfo {author} {\bibfnamefont
  {P.}~\bibnamefont {Zhou}}, \bibinfo {author} {\bibfnamefont {K.}~\bibnamefont
  {Rossnagel}}, \bibinfo {author} {\bibfnamefont {M.}~\bibnamefont {Eckstein}},
  \bibinfo {author} {\bibfnamefont {P.}~\bibnamefont {Werner}},\ and\ \bibinfo
  {author} {\bibfnamefont {U.}~\bibnamefont {Bovensiepen}},\ }\bibfield
  {title} {\bibinfo {title} {Ultrafast doublon dynamics in photoexcited
  $1t$-${\mathrm{tas}}_{2}$},\ }\href
  {https://doi.org/10.1103/PhysRevLett.120.166401} {\bibfield  {journal}
  {\bibinfo  {journal} {Phys. Rev. Lett.}\ }\textbf {\bibinfo {volume} {120}},\
  \bibinfo {pages} {166401} (\bibinfo {year} {2018})}\BibitemShut {NoStop}%
\bibitem [{Note1()}]{Note1}%
  \BibitemOpen
  \bibinfo {note} {A small $b$ captures realistic layers with a large
  super-unit cell}\BibitemShut {NoStop}%
\bibitem [{\citenamefont {Li}\ and\ \citenamefont {Han}(2018)}]{li2018}%
  \BibitemOpen
  \bibfield  {author} {\bibinfo {author} {\bibfnamefont {J.}~\bibnamefont
  {Li}}\ and\ \bibinfo {author} {\bibfnamefont {J.~E.}\ \bibnamefont {Han}},\
  }\bibfield  {title} {\bibinfo {title} {Nonequilibrium excitations and
  transport of dirac electrons in electric-field-driven graphene},\ }\href
  {https://doi.org/10.1103/PhysRevB.97.205412} {\bibfield  {journal} {\bibinfo
  {journal} {Phys. Rev. B}\ }\textbf {\bibinfo {volume} {97}},\ \bibinfo
  {pages} {205412} (\bibinfo {year} {2018})}\BibitemShut {NoStop}%
\bibitem [{\citenamefont {Tsuji}\ \emph {et~al.}(2011)\citenamefont {Tsuji},
  \citenamefont {Oka}, \citenamefont {Werner},\ and\ \citenamefont
  {Aoki}}]{tsuji2011}%
  \BibitemOpen
  \bibfield  {author} {\bibinfo {author} {\bibfnamefont {N.}~\bibnamefont
  {Tsuji}}, \bibinfo {author} {\bibfnamefont {T.}~\bibnamefont {Oka}}, \bibinfo
  {author} {\bibfnamefont {P.}~\bibnamefont {Werner}},\ and\ \bibinfo {author}
  {\bibfnamefont {H.}~\bibnamefont {Aoki}},\ }\bibfield  {title} {\bibinfo
  {title} {Dynamical band flipping in fermionic lattice systems: An
  ac-field-driven change of the interaction from repulsive to attractive},\
  }\href {https://doi.org/10.1103/PhysRevLett.106.236401} {\bibfield  {journal}
  {\bibinfo  {journal} {Phys. Rev. Lett.}\ }\textbf {\bibinfo {volume} {106}},\
  \bibinfo {pages} {236401} (\bibinfo {year} {2011})}\BibitemShut {NoStop}%
\bibitem [{Note2()}]{Note2}%
  \BibitemOpen
  \bibinfo {note} {This may arise either due to a gap exceeding the relevant
  electronic bandwidths, or more generally due to the weakness of
  Coulomb-mediated Auger processes that involve relatively large momentum
  transfers.}\BibitemShut {Stop}%
\bibitem [{Note3()}]{Note3}%
  \BibitemOpen
  \bibinfo {note} {Note that our model contains an even number of electrons per
  unit cell while for 1T-TaS$_2$ it is odd. This leads to a metallic (rather
  than semiconducting) secondary minimum. This difference should, however, not
  change the sliding scenario qualitatively.}\BibitemShut {Stop}%
\bibitem [{\citenamefont {Salzmann}\ \emph {et~al.}(2022)\citenamefont
  {Salzmann}, \citenamefont {Hujala}, \citenamefont {Witteveen}, \citenamefont
  {Hildebrand}, \citenamefont {Berger}, \citenamefont {von Rohr}, \citenamefont
  {Nicholson},\ and\ \citenamefont {Monney}}]{salzmann2022}%
  \BibitemOpen
  \bibfield  {author} {\bibinfo {author} {\bibfnamefont {B.}~\bibnamefont
  {Salzmann}}, \bibinfo {author} {\bibfnamefont {E.}~\bibnamefont {Hujala}},
  \bibinfo {author} {\bibfnamefont {C.}~\bibnamefont {Witteveen}}, \bibinfo
  {author} {\bibfnamefont {B.}~\bibnamefont {Hildebrand}}, \bibinfo {author}
  {\bibfnamefont {H.}~\bibnamefont {Berger}}, \bibinfo {author} {\bibfnamefont
  {F.~O.}\ \bibnamefont {von Rohr}}, \bibinfo {author} {\bibfnamefont {C.~W.}\
  \bibnamefont {Nicholson}},\ and\ \bibinfo {author} {\bibfnamefont
  {C.}~\bibnamefont {Monney}},\ }\bibfield  {title} {\bibinfo {title}
  {Observation of the metallic mosaic phase in 1$ t $-tas $ \_2 $ at
  equilibrium},\ }\href@noop {} {\bibfield  {journal} {\bibinfo  {journal}
  {arXiv preprint arXiv:2209.07945}\ } (\bibinfo {year} {2022})}\BibitemShut
  {NoStop}%
\bibitem [{\citenamefont {Zhang}\ \emph {et~al.}(2022)\citenamefont {Zhang},
  \citenamefont {Gao}, \citenamefont {Cheng}, \citenamefont {Bu}, \citenamefont
  {Wu}, \citenamefont {Fei}, \citenamefont {Zheng}, \citenamefont {Wang},
  \citenamefont {Li}, \citenamefont {Luo} \emph {et~al.}}]{zhang2022}%
  \BibitemOpen
  \bibfield  {author} {\bibinfo {author} {\bibfnamefont {W.}~\bibnamefont
  {Zhang}}, \bibinfo {author} {\bibfnamefont {J.}~\bibnamefont {Gao}}, \bibinfo
  {author} {\bibfnamefont {L.}~\bibnamefont {Cheng}}, \bibinfo {author}
  {\bibfnamefont {K.}~\bibnamefont {Bu}}, \bibinfo {author} {\bibfnamefont
  {Z.}~\bibnamefont {Wu}}, \bibinfo {author} {\bibfnamefont {Y.}~\bibnamefont
  {Fei}}, \bibinfo {author} {\bibfnamefont {Y.}~\bibnamefont {Zheng}}, \bibinfo
  {author} {\bibfnamefont {L.}~\bibnamefont {Wang}}, \bibinfo {author}
  {\bibfnamefont {F.}~\bibnamefont {Li}}, \bibinfo {author} {\bibfnamefont
  {X.}~\bibnamefont {Luo}}, \emph {et~al.},\ }\bibfield  {title} {\bibinfo
  {title} {Visualizing the evolution from mott insulator to anderson insulator
  in ti-doped 1t-tas2},\ }\href@noop {} {\bibfield  {journal} {\bibinfo
  {journal} {npj Quantum Materials}\ }\textbf {\bibinfo {volume} {7}},\
  \bibinfo {pages} {1} (\bibinfo {year} {2022})}\BibitemShut {NoStop}%
\bibitem [{\citenamefont {Werner}\ \emph {et~al.}(2019)\citenamefont {Werner},
  \citenamefont {Eckstein}, \citenamefont {M{\"u}ller},\ and\ \citenamefont
  {Refael}}]{werner2019}%
  \BibitemOpen
  \bibfield  {author} {\bibinfo {author} {\bibfnamefont {P.}~\bibnamefont
  {Werner}}, \bibinfo {author} {\bibfnamefont {M.}~\bibnamefont {Eckstein}},
  \bibinfo {author} {\bibfnamefont {M.}~\bibnamefont {M{\"u}ller}},\ and\
  \bibinfo {author} {\bibfnamefont {G.}~\bibnamefont {Refael}},\ }\bibfield
  {title} {\bibinfo {title} {Light-induced evaporative cooling of holes in the
  hubbard model},\ }\href@noop {} {\bibfield  {journal} {\bibinfo  {journal}
  {Nature Communications}\ }\textbf {\bibinfo {volume} {10}},\ \bibinfo {pages}
  {1} (\bibinfo {year} {2019})}\BibitemShut {NoStop}%
\bibitem [{\citenamefont {Picano}\ \emph
  {et~al.}(2021{\natexlab{a}})\citenamefont {Picano}, \citenamefont {Li},\ and\
  \citenamefont {Eckstein}}]{picano2021prb}%
  \BibitemOpen
  \bibfield  {author} {\bibinfo {author} {\bibfnamefont {A.}~\bibnamefont
  {Picano}}, \bibinfo {author} {\bibfnamefont {J.}~\bibnamefont {Li}},\ and\
  \bibinfo {author} {\bibfnamefont {M.}~\bibnamefont {Eckstein}},\ }\bibfield
  {title} {\bibinfo {title} {Quantum boltzmann equation for strongly correlated
  electrons},\ }\href {https://doi.org/10.1103/PhysRevB.104.085108} {\bibfield
  {journal} {\bibinfo  {journal} {Phys. Rev. B}\ }\textbf {\bibinfo {volume}
  {104}},\ \bibinfo {pages} {085108} (\bibinfo {year}
  {2021}{\natexlab{a}})}\BibitemShut {NoStop}%
\bibitem [{\citenamefont {Picano}\ \emph
  {et~al.}(2021{\natexlab{b}})\citenamefont {Picano}, \citenamefont {Grandi},\
  and\ \citenamefont {Eckstein}}]{picano2021}%
  \BibitemOpen
  \bibfield  {author} {\bibinfo {author} {\bibfnamefont {A.}~\bibnamefont
  {Picano}}, \bibinfo {author} {\bibfnamefont {F.}~\bibnamefont {Grandi}},\
  and\ \bibinfo {author} {\bibfnamefont {M.}~\bibnamefont {Eckstein}},\
  }\bibfield  {title} {\bibinfo {title} {Inhomogeneous disordering at a
  photo-induced charge density wave transition},\ }\href@noop {} {\bibfield
  {journal} {\bibinfo  {journal} {arXiv preprint arXiv:2112.15323}\ } (\bibinfo
  {year} {2021}{\natexlab{b}})}\BibitemShut {NoStop}%
\bibitem [{\citenamefont {Tancogne-Dejean}\ \emph {et~al.}(2020)\citenamefont
  {Tancogne-Dejean}, \citenamefont {Oliveira}, \citenamefont {Andrade},
  \citenamefont {Appel}, \citenamefont {Borca}, \citenamefont {Breton},
  \citenamefont {Buchholz}, \citenamefont {Castro}, \citenamefont {Corni},
  \citenamefont {Correa}, \citenamefont {Giovannini}, \citenamefont {Delgado},
  \citenamefont {Eich}, \citenamefont {Flick}, \citenamefont {Gil},
  \citenamefont {Gomez}, \citenamefont {Helbig}, \citenamefont {Hübener},
  \citenamefont {Jestädt}, \citenamefont {Jornet-Somoza}, \citenamefont
  {Larsen}, \citenamefont {Lebedeva}, \citenamefont {Lüders}, \citenamefont
  {Marques}, \citenamefont {Ohlmann}, \citenamefont {Pipolo}, \citenamefont
  {Rampp}, \citenamefont {Rozzi}, \citenamefont {Strubbe}, \citenamefont
  {Sato}, \citenamefont {Schäfer}, \citenamefont {Theophilou}, \citenamefont
  {Welden},\ and\ \citenamefont {Rubio}}]{tancogne2020}%
  \BibitemOpen
  \bibfield  {author} {\bibinfo {author} {\bibfnamefont {N.}~\bibnamefont
  {Tancogne-Dejean}}, \bibinfo {author} {\bibfnamefont {M.~J.~T.}\ \bibnamefont
  {Oliveira}}, \bibinfo {author} {\bibfnamefont {X.}~\bibnamefont {Andrade}},
  \bibinfo {author} {\bibfnamefont {H.}~\bibnamefont {Appel}}, \bibinfo
  {author} {\bibfnamefont {C.~H.}\ \bibnamefont {Borca}}, \bibinfo {author}
  {\bibfnamefont {G.~L.}\ \bibnamefont {Breton}}, \bibinfo {author}
  {\bibfnamefont {F.}~\bibnamefont {Buchholz}}, \bibinfo {author}
  {\bibfnamefont {A.}~\bibnamefont {Castro}}, \bibinfo {author} {\bibfnamefont
  {S.}~\bibnamefont {Corni}}, \bibinfo {author} {\bibfnamefont {A.~A.}\
  \bibnamefont {Correa}}, \bibinfo {author} {\bibfnamefont {U.~D.}\
  \bibnamefont {Giovannini}}, \bibinfo {author} {\bibfnamefont
  {A.}~\bibnamefont {Delgado}}, \bibinfo {author} {\bibfnamefont {F.~G.}\
  \bibnamefont {Eich}}, \bibinfo {author} {\bibfnamefont {J.}~\bibnamefont
  {Flick}}, \bibinfo {author} {\bibfnamefont {G.}~\bibnamefont {Gil}}, \bibinfo
  {author} {\bibfnamefont {A.}~\bibnamefont {Gomez}}, \bibinfo {author}
  {\bibfnamefont {N.}~\bibnamefont {Helbig}}, \bibinfo {author} {\bibfnamefont
  {H.}~\bibnamefont {Hübener}}, \bibinfo {author} {\bibfnamefont
  {R.}~\bibnamefont {Jestädt}}, \bibinfo {author} {\bibfnamefont
  {J.}~\bibnamefont {Jornet-Somoza}}, \bibinfo {author} {\bibfnamefont {A.~H.}\
  \bibnamefont {Larsen}}, \bibinfo {author} {\bibfnamefont {I.~V.}\
  \bibnamefont {Lebedeva}}, \bibinfo {author} {\bibfnamefont {M.}~\bibnamefont
  {Lüders}}, \bibinfo {author} {\bibfnamefont {M.~A.~L.}\ \bibnamefont
  {Marques}}, \bibinfo {author} {\bibfnamefont {S.~T.}\ \bibnamefont
  {Ohlmann}}, \bibinfo {author} {\bibfnamefont {S.}~\bibnamefont {Pipolo}},
  \bibinfo {author} {\bibfnamefont {M.}~\bibnamefont {Rampp}}, \bibinfo
  {author} {\bibfnamefont {C.~A.}\ \bibnamefont {Rozzi}}, \bibinfo {author}
  {\bibfnamefont {D.~A.}\ \bibnamefont {Strubbe}}, \bibinfo {author}
  {\bibfnamefont {S.~A.}\ \bibnamefont {Sato}}, \bibinfo {author}
  {\bibfnamefont {C.}~\bibnamefont {Schäfer}}, \bibinfo {author}
  {\bibfnamefont {I.}~\bibnamefont {Theophilou}}, \bibinfo {author}
  {\bibfnamefont {A.}~\bibnamefont {Welden}},\ and\ \bibinfo {author}
  {\bibfnamefont {A.}~\bibnamefont {Rubio}},\ }\bibfield  {title} {\bibinfo
  {title} {Octopus, a computational framework for exploring light-driven
  phenomena and quantum dynamics in extended and finite systems},\ }\href@noop
  {} {\bibfield  {journal} {\bibinfo  {journal} {The Journal of Chemical
  Physics}\ }\textbf {\bibinfo {volume} {152}},\ \bibinfo {pages} {124119}
  (\bibinfo {year} {2020})}\BibitemShut {NoStop}%
\bibitem [{\citenamefont {Garavelli}\ \emph {et~al.}(1997)\citenamefont
  {Garavelli}, \citenamefont {Celani}, \citenamefont {Bernardi}, \citenamefont
  {Robb},\ and\ \citenamefont {Olivucci}}]{garavelli1997}%
  \BibitemOpen
  \bibfield  {author} {\bibinfo {author} {\bibfnamefont {M.}~\bibnamefont
  {Garavelli}}, \bibinfo {author} {\bibfnamefont {P.}~\bibnamefont {Celani}},
  \bibinfo {author} {\bibfnamefont {F.}~\bibnamefont {Bernardi}}, \bibinfo
  {author} {\bibfnamefont {M.}~\bibnamefont {Robb}},\ and\ \bibinfo {author}
  {\bibfnamefont {M.}~\bibnamefont {Olivucci}},\ }\bibfield  {title} {\bibinfo
  {title} {The c5h6nh2+ protonated shiff base: an ab initio minimal model for
  retinal photoisomerization},\ }\href@noop {} {\bibfield  {journal} {\bibinfo
  {journal} {Journal of the American Chemical Society}\ }\textbf {\bibinfo
  {volume} {119}},\ \bibinfo {pages} {6891} (\bibinfo {year}
  {1997})}\BibitemShut {NoStop}%
\bibitem [{\citenamefont {Martinez}(2010)}]{Martinez2010}%
  \BibitemOpen
  \bibfield  {author} {\bibinfo {author} {\bibfnamefont {T.}~\bibnamefont
  {Martinez}},\ }\bibfield  {title} {\bibinfo {title} {Seaming is believing},\
  }\href@noop {} {\bibfield  {journal} {\bibinfo  {journal} {Nature}\ }\textbf
  {\bibinfo {volume} {467}},\ \bibinfo {pages} {412–} (\bibinfo {year}
  {2010})}\BibitemShut {NoStop}%
\bibitem [{Note4()}]{Note4}%
  \BibitemOpen
  \bibinfo {note} {Usually there are additional directional factors taking into
  account the orientation of chemical bonds. We neglect such factors for
  simplicity.}\BibitemShut {Stop}%
\end{thebibliography}%
\end{document}